\documentclass{elsart}
\usepackage{graphicx,amssymb,times}
\journal{NIM}
\usepackage{graphics}
\include{graphics}
      \usepackage{amssymb}
         \usepackage{lineno}
      \begin{document}
      \begin{frontmatter}
\corauth[cor1]{Corresponding author: mkkoul@barc.gov.in }
\corauth{Tel: +91-022-25591792 / 25515209}
\title{ {Simulation studies for optimizing the trigger generation  criteria  
for the TACTIC telescope}\\
}
\author{M.K. Koul$^*$, A.K. Tickoo, V.K. Dhar, K. Venugopal, K. Chanchalani,}
\author{ R.C. Rannot, K.K.Yadav, P. Chandra, M. Kothari, R. Koul}

\address{\it Astrophysical Sciences Division, Bhabha Atomic Research Centre, Mumbai-400085, India.\\}
\begin{abstract}
In this paper,  we present  the results  of  Monte Carlo  simulations  of   $\gamma$-ray and 
cosmic  ray  proton induced  extensive  air  showers  as  detected  by the TACTIC   atmospheric  Cherenkov
  imaging telescope  for  optimizing   its  trigger  field of view  and  topological  trigger  generation  
scheme.  The  simulation study   has   been  carried out  at several  zenith angles.  The    topological  
trigger  generation  uses   a coincidence  of  2 or  3 nearest  neighbour  pixels  for producing   an  event
  trigger.    The  results   of this  study  suggest   that  a trigger  field  of   11$\times$11  pixels
 ($\sim$ $3.4^o$ $\times$ $3.4^o$)  is quite  optimum   for  achieving   maximum
effective collection    area  for    $\gamma$-rays  from a point  source.  With regard  to  optimization
  of  topological  trigger  generation,   it  is found  that   both   2  or  3  nearest  neighbour  pixels
   yield   nearly similar  results   
up to a zenith  angle of  ~$25^o$   with a    threshold  energy of   $\sim$ 1.5 TeV  for  $\gamma$-rays.
Beyond  zenith  angle of  ~$25^o$, the   results  suggest that  a  2-pixel   nearest  neighbour  trigger 
  should be preferred.   
Comparison of the  simulated  integral  rates    has also been  made  with   corresponding   measured
   values  for   validating  the  predictions  of  the   Monte Carlo  simulations,   especially    the  
effective  collection area,   so that  energy  spectra  of  sources  (or  flux upper limits in case of  no 
detection)  can  be  determined  reliably.  Reasonably good  matching   of the   measured  trigger  rates  
(on  the basis   of  $\sim$ 207  hours of  data  collected  with  the telescope  in  NN-2  and  NN-3   
trigger configurations)   with  that obtained  from simulations  reassures  that the procedure followed by 
us  in  estimating    the  threshold  energy  and  detection rates   is  quite  reliable.
                   
\end{abstract}
\end{frontmatter}
\section{Introduction}
\label{1}
Detection  of  cosmic  $\gamma$-rays  in the  very  high energy range ( $\sim$ 100 GeV-50 TeV) using  
ground-based atmospheric 
Cherenkov  technique  [1-2] allows  a detailed  study of  the nonthermal universe and  probes  many  
phenomena  of  fundamental
astrophysical interest  including  the acceleration and emission mechanisms  in galactic and 
extragalactic sources, physics of  pulsars  and  supernova remnants, study of  extragalactic  diffuse  
background etc.
The  recent  success  of  the  field   has mainly   resulted  from the  development  of the  Cherenkov   
imaging  telescopes 
which  allows  efficient   separation  of  photon  induced  showers  from the hadron  background.  The 
power  of this  technique  has  been  convincingly  demonstrated by  several   modern  atmospheric   
Cherenkov  telescopes  ( e.g.  MAGIC [3], VERITAS [4], HESS [5] and CANGAROO [6])   and   there  is   now 
a general  consensus    that   stereoscopic   arrays    provide   the  most  promising  avenue  for future
  observations.     
\par
A  typical  imaging  atmospheric  Cherenkov  telescope  consists  of  a  light  collector  and a  
close-packed array  of  fast photomultiplier  tubes ( also called  the imaging camera  with  individual  
tubes  as its pixels). The imaging  camera  records   the spatial distribution of  photons in the image 
(called   the Cherenkov image) and by exploiting   the differences in  the  images produced  by  $\gamma$-ray
 and  cosmic-ray initiated  showers,  caused by  the  physical processes responsible for  their development
  in the atmosphere,  it  becomes possible to effectively  segregate the two event  species   on  a  statistical  basis  
  with a high 
degree of accuracy.  Simulation  studies of  atmospheric  electromagnetic ($\gamma$-ray induced)
  and  hadronic ( proton induced)  cascades  [7,8]    have shown   that    there  
exist  subtle  differences  in the image  features ( shape, size and   orientation) of  these  event types.
   The  image  of   an  off-axis  $\gamma$-ray   
shower  is  relatively    more  compact  and  is  elliptical  in shape, with the major   axis  of the 
ellipse   oriented  towards  the  camera  centre (i.e. the source position).  In the  case of  off-axis  
cosmic  ray shower,  on the contrary,  the  image is  comparatively bigger  and irregular, besides  having
  no preferred  orientation (because of  isotropic  nature of  cosmic rays).
Remarkably indeed, these  predictions  have been thoroughly matched by observations  of various groups
   with   great success  and  
it  is  primarily   because  of  the success of these  experiments  that the  field  of  ground-based
   $\gamma$-ray   astronomy  today  boasts a  catalog  of  more  than  a   hundred  $\gamma$-ray  sources
 [9].
\par
In  an atmospheric  Cherenkov  imaging  telescope,   there  are  several   crucial parameters which  need  to be
  fine-tuned    for  maximizing the   performance  of  the  system.   Assuming   that  we  are  observing
  a  point  $\gamma$-ray source  with  the  telescope  axis  directly  pointing  towards  it,   these  are :
   (a) the point spread  function of the   light  collector,
   (b) the pixel size  of the imaging  camera, (c)  trigger field of view  and  (d) trigger generation scheme.  
The optimum  trigger  field  of  view  of  an  atmospheric  Cherenkov  imaging  telescope   needs  to be   
chosen  in such a manner    that  the  ratio of  $\gamma$-rays  to  the  cosmic  ray background  is  
maximum.   Thus,  for   an  imaging  camera  with  a  large field of  view, it is  generally advantageous 
to exclude  the outer pixels  from  the trigger  generation.  These  pixels  only  help to  properly  
reconstruct  the image parameters  and  do  not  increase  the   $\gamma$-ray rate.   The  trigger in  
an  atmospheric  Cherenkov  imaging  telescope  is  generated  by   demanding  a  coincidence  between 
a number  of  PMTs, with  the  condition  that  pixels  participating  in the  trigger should  have  a   
signal  exceeding  some   threshold  q$_{0}$.  The  conventionally  used   multi-fold ( 2 to 4) coincidence 
approach 
necessitates the operation of the telescope 
at a relatively higher threshold energy in order to keep the chance coincidence rate  at
a reasonable value.  An improved trigger generation scheme involving two or three nearest neighbour pixels 
can be used instead  
for  lowering  the  threshold  energy  of the  telescope.  The  main  aim  of this  work  is  to  use  
Monte  Carlo simulations for  optimizing  the  above  mentioned  parameters  for the  TACTIC  telescope.             
        
\section{ TACTIC imaging telescope}
\label{}
The TACTIC (TeV Atmospheric Cherenkov Telescope with Imaging  Camera) $\gamma$-ray telescope [10] has been 
in operation  at Mt. Abu ( 24.6$^\circ$ N,  72.7$^\circ$ E, 1300m  asl), India, for the last several years
  to  study TeV gamma ray emission  from celestial sources.  The  telescope   uses a  tessellated  
light-collector of  area  $\sim$ 9.5m$^2$ which is  capable of tracking a celestial source across the sky.
  The telescope deploys a  349-pixel  imaging camera,   with a uniform pixel size   of 
$\sim$ $0.31^o$ and a $\sim$ $5.9^o$$\times$$5.9^o$ field-of-view,   to record  atmospheric 
Cherenkov events produced by an incoming cosmic-ray particle or a $\gamma$-ray photon with an energy above
  $\sim$1TeV. The  TACTIC light-collector  uses 34   front-face aluminium-coated,  glass spherical mirrors
 of 60 cm diameter  each  with  a focal length $\sim$ 400cm 
and reflection coefficient $>$80$\%$ at a wavelength of $\sim$ 400nm. The  point-spread  function  has a  
FWHM of $\sim$ 0.185$^0$ ($\equiv$12.5mm)  and  D$_{90}$  $\sim$ 0.34$^0$ ($\equiv$22.8mm).   Here, D$_{90}$
  is   defined as the diameter of  circle, concentric  with the centroid of the image, within which 90$\%$  
of reflected rays lie.  The value of  D$_{90}$  $\sim$ 0.29$^0$ ($\equiv$19.3mm),  predicted on the basis of 
the simulation  for an incidence angle of 0$^0$,  matches  reasonably well with the measured  value 
mentioned above. Other details regarding the   ray-tracing simulation procedure  and comparison of the 
measured  point-spread function of the TACTIC  light collector with the simulated performance of ideal 
Davies-Cotton and paraboloid designs  are discussed in [11]. 
The  photographs of the TACTIC  imaging  telescope and its  back-end  signal processing electronics are  
shown in Fig.1 
\begin{figure}[h]
\centering
\includegraphics*[width=1.0\textwidth,angle=0,clip] {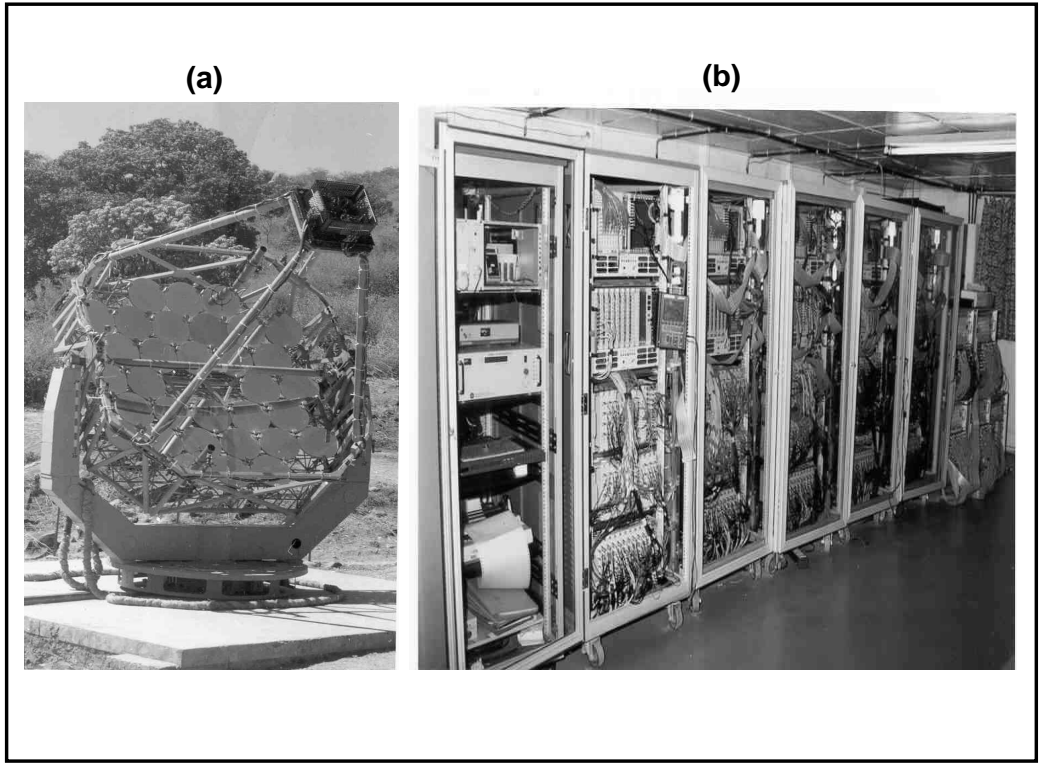}   
\caption{\label{fig ---} (a) Photograph of the 349-pixel TACTIC  imaging  telescope (b) Photograph of 
back-end  signal processing electronics used in the telescope.}
\end{figure}
\par

The  imaging  camera uses 19mm diameter  photomultiplier tubes  (ETL-9083 UVB). The bialkali  photocathode 
has a maximum quantum efficiency of $\sim$ 27$\%$   and   for this   PMT  the   maximum  quantum efficiency 
is  usually   reached  in the wavelength  range   $\sim$ 340 -360 nm.   In order  to  ensure  better  light collection 
efficiency and reduction in the background  light falling  on the photomultipliers   we    use   Compound  
Parabolic Concentrator (CPC) as  light guides  on the photomultipliers.  
The  CPCs  used  in the  TACTIC   imaging  camera   have   the   following  specifications :  entry aperture $\sim$21.0 mm; exit aperture $\sim$15.0 mm; and height $\sim$17.6 mm.  The  light collection  efficiency   of the CPCs  is  estimated  to be  
of $\sim$65$\%$   by  using   ray  tracing  simulations  and  includes   both  the geometrical collection efficiency and the 
reflectivity  of the surface. The  light collection  efficiency mentioned  above  refers   to  those   photons  which undergo reflection  
after falling  on the  inside surface of  the  CPCs.  The data acquisition and control system of the telescope [12] has been designed around a network of PCs 
running the QNX (version 4.25) real-time operating system. The triggered events are digitized by CAMAC 
based 12-bit Charge to Digital Converters (CDC) which have a full scale range of 600 pC. The relative  gain 
of the photomultiplier tubes  is monitored   by repeatedly  flashing   a blue LED, placed  at a distance 
of $\sim$1.5m from the camera. Whenever the single channel rate of any two  or  more pixels in the trigger
 region goes outside the preset operational band, it is automatically restored to within the prescribed
 range by appropriately adjusting the high voltage of the pixels [13].
\par
The imaging  camera  uses  a programmable  topological trigger  [14]  which  can pick up  events  with  a
 variety of
trigger configurations. As  the trigger  scheme is not  hard wired,  a number of coincidence  trigger 
options 
( e.g Nearest Neighbour Pairs,  Nearest Neighbour  Non-collinear Triplets and Nearest Neighbour  
Non-Collinear  Quadruplets)  can be generated under software control.  The  number of  pixels  participating
 in the trigger  can  be  varied   from  7$\times$7  pixels  to   15$\times$16 pixels.  Apart from 
generating  the  prompt trigger with  a coincidence  gate width  of  $\sim$18ns, the trigger generator has
 a provision for producing a chance coincidence output based on  $^{12}$$C_{2}$  combinations from various
 groups of closely spaced 12 channels. It is  worth mentioning  here  that   there  are  several  reasons   which  
compel  us  to  use  a  coincidence  gate width  of  $\sim$18 ns   instead  of   an  optimal gate width  $\leq$ 5 ns.    
The  reasons  are  the following: (a)  variation   in  the   transit  time   of  the  PMTs     because   of   being    operated  at   different   high voltage values, (b)  variation   in  the   propagation    delay   of  50m long    RG58  cables  because of   procurement  of     the  cables  in different   batches, (c)  degradation  in  the   rise  time   of  the  PMT  pulses    when    the     signal  is  transmitted     through   a    50m  long    RG58  cable   and   the   subsequent    'time-walk'   of  the   leading  edge  discriminators  which   is  dependent  on  the    amplitude   of  the   PMT  output  signal  and  (d)  variation  in the   access  time   of     static  RAMS ( Toshiba make TC55B417)  used   in   the     trigger  generator   and  difference  of  $\sim$ 4ns    in the cascading  and  the non-cascading   routes  [14].  Feasibility   of    using   programmable   digital delay  generators  for  fine  tuning   the signal  propagation  time in all  the channels  is also  being   investigated  so  that  the    coincidence    gate   can   be  reduced   to    
$<$ 10ns.  Other details  regarding the  design, implementation and performance
evaluation  of the programmable topological trigger generator  for the 349-pixel imaging camera of the
TACTIC  telescope  are discussed in [14]. 
\par
A user friendly in-house developed  tracking  system  software provides  an independent movement  for the 
zenithal  and azimuthal  axes  so that a matching  between the telescope  pointing  direction  and  the 
source  direction is obtained  with an accuracy  better than  $\pm$2 arc-min. The tracking accuracy  of 
the telescope  is also checked  on a regular basis  with so called 'point runs', where a reasonably  
bright  star,    having  a  declination  close  to that of the  candidate $\gamma$-ray source is  
tracked continuously  for about 5 hours.  The  point run calibration data  (corrected zenith and 
azimuth angle  of the telescope  when the star image is centered)  are  then incorporated in the 
telescope drive system software so that appropriate corrections can be  applied directly  in real time  
while tracking  a candidate $\gamma$-ray source. 
Other relevant details  of the instrumentation  aspects of the telescope  
are  discussed [10]. 
\section {Importance of using neighbouring pixels for trigger generation and TACTIC trigger generator design}
\label{}
The conventional  approach  of generating  an  event  trigger  in  atmospheric  Cherenkov  imaging 
telescopes is  to use  a majority  type   coincidence   with  trigger  multiplicity  of   2 to 4. The  
optimum  value  of the trigger  multiplicity  is  decided 
by several  factors  like   observatory altitude,  primary  energy and zenith angle  dependent  intensity  distribution of 
the Cherenkov  photons  in the image plane  and   the pixel size.    
However, in order to keep the chance coincidence rate  at 
a reasonable value (usually few percent of prompt coincidence rate), the  use of  a majority type   
coincidence approach ( i.e.  any $\it m$ present out of total $\it n$ detector channels, denoted  by  
any-m,n  here onwards ), in  generating the trigger,  
leads  to a relatively  higher threshold energy. The  reason  for this   being  the  presence of 
   $^nC_m$  factor  in m-fold chance coincidence rate ($\it f_{any-m,n}$) formula which is given by
\begin{equation}
\it f_{any-m,n} \approx C_{any-m,n}  m  R_{pixel}(q_0)^m \tau_{c}^{m-1}= m  \frac{ n!} { m!(n-m)!}  R_{pixel}(q_0)^m \tau_{c}^{m-1} \
\end{equation}
where   $C_{any-m,n}$= $^nC_m$= $ n!/ (m!(n-m)!)$ is  the number of  all possible  m-pixel  combinations,   
 $\it R_{pixel}(q_0)$  is the shot-noise generated single channel rate  passing  the  discriminator  set 
at a threshold  of  $\it q_0$  (in  units of photoelectrons), $\it m$ is the number of 
detector channels which  fire coincidently within a coincidence resolving time $\tau_{c}$,  $\it n$ is the 
total number 
of detector channels which can  participate in producing the trigger. It has  been assumed   in  the above  
formula   that  the contribution  from  higher  multiplicity  triggers  (i.e.  due  to coincidence  of  
$>m$  channels)  is negligible   as  compared  to  the  m-fold  chance coincidence rate  and   this can  
be satisfied  easily as long as  $R_{pixel}(q_0)\tau_{c}<<$1.  Keeping  in view  the compact  nature of 
$\gamma$-ray  images,  an elegant  way to reduce the  chance coincidence rate is  to  limit  the  number 
of $\it m$-fold  combinations   by  using  topological  constraints  in the trigger (such as  strict 
adjacency  of triggering  pixels  or  demanding  that the pixels  should  lie within camera sectors). 
If $C_{NN-m,n}$  is  the number  of  m-fold  nearest  neighbour  combinations the corresponding 
chance coincidence rate ($\it f_{NN-m,n}$)  is   given by : 
\begin{equation}
\it f_{NN-m,n}  \approx  C_{NN-m,n}  m  R_{pixel}(q_0)^m \tau_{c}^{m-1} \
\end{equation}
Since,  the number of nearest neighbour  m-fold   combinations ( $C_{NN-m,n}$)   is always  less than  the 
corresponding   number of combinations   in  a simple  majority  type coincidence ( $C_{any-m,n}$),   there 
is   a significant  reduction in the chance  coincidence  rate   when  nearest neighbour  m-fold   trigger
  generation  is used.  Thus,    use  of  a nearest neighbour  m-fold   based  trigger  generation  allows
  the  individual  pixels  to operate  at a relatively lower  threshold   which eventually  leads  to 
lowering  the  threshold   energy of the telescope. 
\par
An improved trigger generation scheme, based on the above  concept  was proposed
by us  [15] for the Whipple system and  a reduction in threshold energy by about 20$\%$  is   achieved by 
the  group when   a  2 adjacent /331  trigger condition instead of  the conventional $^{331}C_2$ trigger 
condition  is  used [16]. 
This has resulted in $\sim$40$\%$ increase in the number of $\gamma$-rays detected from the Crab Nebula [16].
The usefulness of the nearest-neighbour  trigger generation scheme  in  achieving   the lowest possible 
energy threshold  has  been  demonstrated  by    other  telescopes  also including  VERITAS, HESS and 
MAGIC [17,18].  
The  TACTIC imaging  telescope   also  employs  a  programmable  topological  trigger  generator   with  
2( or 3)  nearest neighbour pixels  participating  in the trigger  and   the  main   aim  of this  work  
is  to  optimize   the overall  trigger  generation   scheme, both  with respect  to  trigger multiplicity 
(i.e.  NN-2  versus NN-3) as well as  number of   pixels  participating   in the trigger.  Thus, the  work
involves   comparing the  performance  of  NN-2  and  NN-3  trigger  schemes  with  number of  pixels   
participating   in the  trigger  varying  from  49 to 240. 
\par
The  number of   2-fold  and  3-fold  combinations  for  majority type   and  Nearest-Neighbour 
coincidence    for different  numbers  of pixels  participating  in the  trigger  is  given in Table.1  
\begin{table}[h]
\centering
\caption{ The  number of  2-fold  and  3-fold  combinations  for  majority   and  Nearest-Neighbour  
coincidence  trigger schemes. }
\begin{tabular}{|c|c|c|c|c|c|}
\hline
\hline
No. of Trigger   &   Field of  &  \multicolumn{4}{c|}{Number of Possible combinations} \\
\cline{3-6}
pixels (n) &  View  &\multicolumn{2}{c|}{2-fold}&\multicolumn{2}{c|}{3-fold}\\
\hline
&  &    $ C_{any-2,n}$ & $C_{NN-2,n}$ &  $C_{any-3,n}$ &  $C_{NN-3,n}$ \\

\hline
\hline
7$\times$7 (= 49) & $2.17^o\times2.17^o$ & 1176 & 156 & 18424 & 624 \\
           
\hline
9$\times$9 (= 81) & $2.79^o\times2.79^o$ & 3240 & 272 & 85320 & 1152 \\

\hline
11$\times$11 (= 121) & $3.41^o\times3.41^o$ & 7260 & 420 & 287980 & 1840 \\      

\hline
13$\times$13 (= 169) & $4.03^o\times4.03^o$ & 14196 & 600 & 790244 & 2688 \\ 
       
\hline
16$\times$15 (= 240) & $4.96^o\times4.65^o$ & 28680 & 869 & 2275280 & 3968 \\ 

\hline
\hline

\end{tabular}
\end{table}
These   combinations   have  been  worked  out  by considering  the  actual  layout   of  the   pixels  
used  in the TACTIC  imaging  camera  ( Fig.2a).  Different    trigger  regions  with  49,81,121,169 and  
240  pixels   considered  in this  work are also  indicated in this  figure.  Defining  
$\eta_{m,n}= C_{NN-m,n}/C_{any-m,n}$  as the  reduction in the  chance  coincidence  rate,  Fig. 2b  
shows  the   variation of   $\eta_{2,n}$  and   $\eta_{3,n}$ as a function of n.  
\begin{figure}[h]
\includegraphics*[width=0.9\textwidth,angle=0,clip]{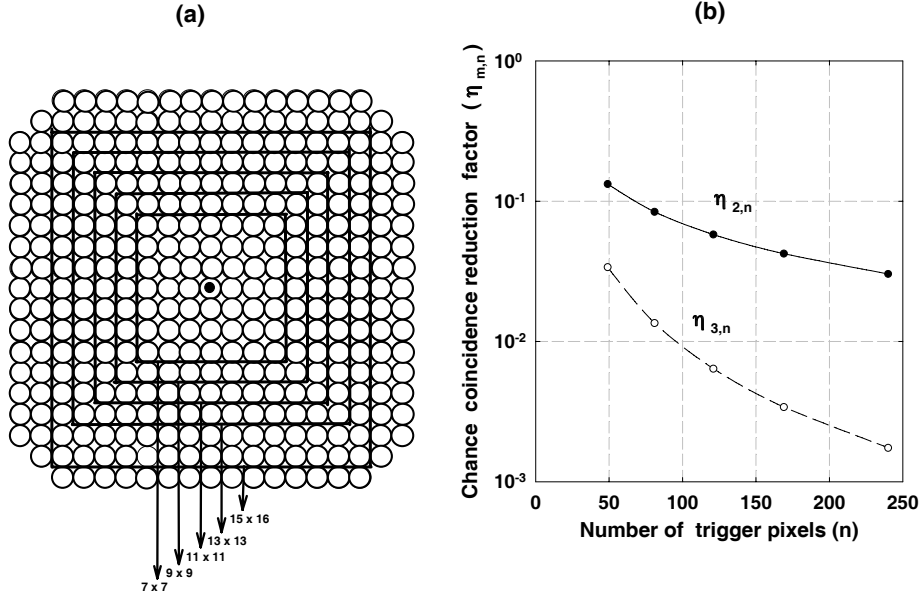}
\caption{ (a) Layout  of the  349-pixels  of the  TACTIC  imaging  camera. Different    trigger  regions
  with  49, 81, 121, 169  and  240  pixels   considered  in this  work are also  indicated in the figure. 
(b) Variation of   $\eta_{2,n}$  and   $\eta_{3,n}$ as a function of number  of  pixels  in the   
trigger region.}
\end{figure}
With  $\eta_{2,n}\sim 4.822$ $n^{-0.923}$   and    $\eta_{3,n}\sim 44.182$ $n^{-1.843}$, as  seen from 
Fig.2b,  it is obvious   that  using  a nearest  neighbour   trigger  approach  offers  an  elegant  way  
to    reduce  the chance  coincidence  rate. Allowing   only  certain m-pixel combinations to participate  
in the trigger  also  helps  in  removing   a pixels (or certain pixels)  with excessive   single  pixel 
rate  from the  trigger  without  making   hardware  changes.  A very  common  reason  for the excessive    
single  pixel  rate   is    due  to presence  of  relatively  bright  stars  in the   field  of view of the 
telescope ( e.g. a star  like  $\zeta$-Tauri    when   Crab  nebula  is  being  tracked  by  a telescope).   
\par
Regarding  the   design  aspects of  the   TACTIC  trigger  generator,  while   a detailed  description of  
the topological trigger  used in the TACTIC  telescope  can be found in  [14], we will  only present  
here  some of  its  key  design features.  The  programmable topological trigger  used  in the TACTIC  
telescope   is   based  on  16k x 4 bit  fast static RAMs   and 
has   a  coincidence  resolving  time of  $\sim$18ns.
As  the trigger  scheme is not  hard wired,  a number of coincidence  trigger options 
( e.g  NN-2,  NN-3  etc.) can be generated under software control.   Given  the   rectangular  layout  
of the  TACTIC  imaging  camera  pixels   we  use  a maximum  of   240  inner  pixels  for trigger  
generation.  Since  each  16k x 4 bit  RAM  can  accommodate  only  16  pixels, we  have  divided   
the  trigger pixels into 20 cells of 12 pixels each in the form of a 3$\times$4  matrix.  The outputs from  
each of the four independently operating trigger generator modules are then collated in an Event Handler 
module  which generates the 22ns duration gate pulse and interrupts the Data Acquisition System (DAS)
for reading the charge ADC data from all the 349 pixels.
\section {Behaviour  of  single pixel rate  and effect of  afterpulsing  in PMTs}
\label{}
It  is  a very well known fact  that  the trigger  threshold  of an atmospheric  Cherenkov telescope  is  
dictated
by  afterpulsing in PMTs  instead  of  fluctuations  in the number of photoelectrons   due  to the light 
of  night sky (LONS) background alone  [19,20]. The  main aim  here is  to  determine  the  single  pixel  
threshold  in a realistic  manner  so 
that the  same  can be  used  in  Monte Carlo  simulations  for  reliably  predicting  the  telescope   
performance ( e.g.  trigger efficiency  as a  function of energy, differential  rates etc.) when  it   
responds  to   $\gamma$-ray and  cosmic ray  showers. 
The  phenomenon of afterpulsing  in  PMTs leads  to generation  of   large amplitude  pulses  (well  in 
excess  of  10 photoelectrons) at  a non-negligible rate. Thus,  it is the  single pixel rate and the 
resulting  maximum  permissible  chance coincidence  rate  which  decides the   minimal  discriminator  
threshold  and  ultimately  the  energy threshold  of the telescope. While  the phenomenon   of 
afterpulsing  (also known as  ion feedback) has  been  investigated thoroughly by a number of  workers,  
we will only  discuss  here,  for   the sake of completeness,   some  important  consequences of  this 
effect. Afterpulsing   is a  result of  ionization  of a molecule or an atom of the residual gas 
inside the  PMT (or  adsorption at  dynode  surface material)  by  photoelectron  and   subsequent  
release of  many photoelectrons  from the photocathode  when the  positively  charged  ion  hits  the  
photocathode.
The relevance of afterpulses depends on how the PMTs are used and they can largely be inconsequential if 
the main pulse can be gated. Since  the  PMTs  used  in atmospheric  Cherenkov imaging cameras are 
operated in a self-trigger mode (i.e.  the PMTs are live all the time), afterpulses  pose a serious 
problem. 
\par
Before  presenting  the measured  behaviour  of  single pixel rate as a function  of threshold  level, 
it  is  worth    estimating     the  single pixel rate ( $R_{pixel}$)   as  a  function of the  threshold  
level  on  the  basis of Poissonian  fluctuations  in the  number of photoelectrons.  Assuming   that  
the  individual  LONS  photons (and hence  the photoelectrons) are uncorrelated   and  the  discriminator  
threshold  is  set  at  $q_0$  photoelectrons,   $R_{pixel}(q_0)$  can be estimated by the  following  
formula :   
 \begin{equation}
 R_{pixel}(q_0) \approx   R_{pe}   \frac{  (R_{pe} \tau_s)^{q_0-1}} { (q_0-1)!  }  \exp ({-R_{pe} \tau_s})  \\
\end{equation}
where $\tau_{s}$  is the characteristic  shaping  time  of  the single  photoelectron  response of the 
PMT ( approximated  as a  rectangular  pulse);  $R_{pe}$ is  the  single photoelectron rate  at which  
photoelectrons  are generated;  $R_{pixel}$  is trigger rate  such that,   within  a time   $\tau_{s}$,    
there  are  $q_0$ photoelectrons.  It has  been also assumed  in  equation 3  that 
$R_{pixel}(q_0)$ $\ll$  $R_{pe}$  and  contribution from   $>(q_0)$  electrons  is  negligible  to  the  
trigger  rate.   The  single photoelectron  rate ($R_{pe}$),
for  a  given   light intensity,   can  be  obtained  by monitoring  the  average PMT  current  ($I_a$) 
and  the  PMT gain 
($G_{PMT}$) and  then  using  the   following  formula: 
\begin{equation} 
 R_{pe}=  \frac{ I_a}{ e G_{PMT}} \\
\end{equation}  
where   e is  the electron  charge. An  estimate   of the  threshold   level ($(q_0)$),   on the  basis  
of  the  equation 3, can now  be readily  worked  out  for a  particular  telescope  system.  Using  a   
light of night sky background  intensity   $\sim$ 2 $\times$  10$^{12}$ photons  $m^{-2}s^{-1}sr^{-1}$ 
in the  wavelength  range   $\sim$320-650nm  for  the Mount Abu observatory  and  other  relevant  
hardware  specifications  of  the  TACTIC  telescope ( e.g., light collection area, pixel solid angle,  
quantum efficiency  of the PMT  etc. ) we  find   that  average  number  of  photoelectrons  per 10ns   
time  interval  is  $\sim$0.20.  With  a  
 single  photoelectron rate  of  $\sim$20 MHz,  on  using  equation 3,   it turns  out   that  the   PMT  
trigger  rate    should be   $\sim$ 1 kHz  and
 $\sim$2.3 $\times 10^{-3}Hz$ at  threshold  levels  of  5 and 10 photoelectrons, respectively.    Actual 
measurements  of the PMT  rate  as a function   of threshold, however,  yield  much  higher    values  
because  of   afterpulsing.   Realizing   that  the  discriminator  threshold  is a very  crucial  input  
required  for  performing   Monte Carlo simulations  in a  realistic  manner   it  becomes  necessary  to 
take  actual  trigger  rate  measurements.  Accordingly,  we have  also  carried  out   these  
measurements   for several PMTs  and the results  of this   study  are discussed  below.
\par
For determining   the  afterpulsing   behaviour,  
laboratory  measurements were  carried out  for  several PMTs  by   simulating   the  LONS   with    a  
LED.  The intensity  of  the  LED  was   set  in such  a  manner, that  the  resulting  PMT anode 
current  is  equivalent  to  the  value  which  the  PMT  would  produce during  actual observations. A  
summary   of the results   obtained  from these  measurements   is  shown in Fig.3.   The integral  count   
rate  as  a  function of  threshold for  1 PMT    using  a  LONS  background  $\sim 2.0\times 10^{12}$ 
photons $m^{-2}s^{-1}sr^{-1}$  is  shown  in Fig.3a ( marked  as  10.3 $\mu$A). 
\begin{figure}[h]
\centering
\includegraphics*[width=1.0\textwidth,angle=0,clip]{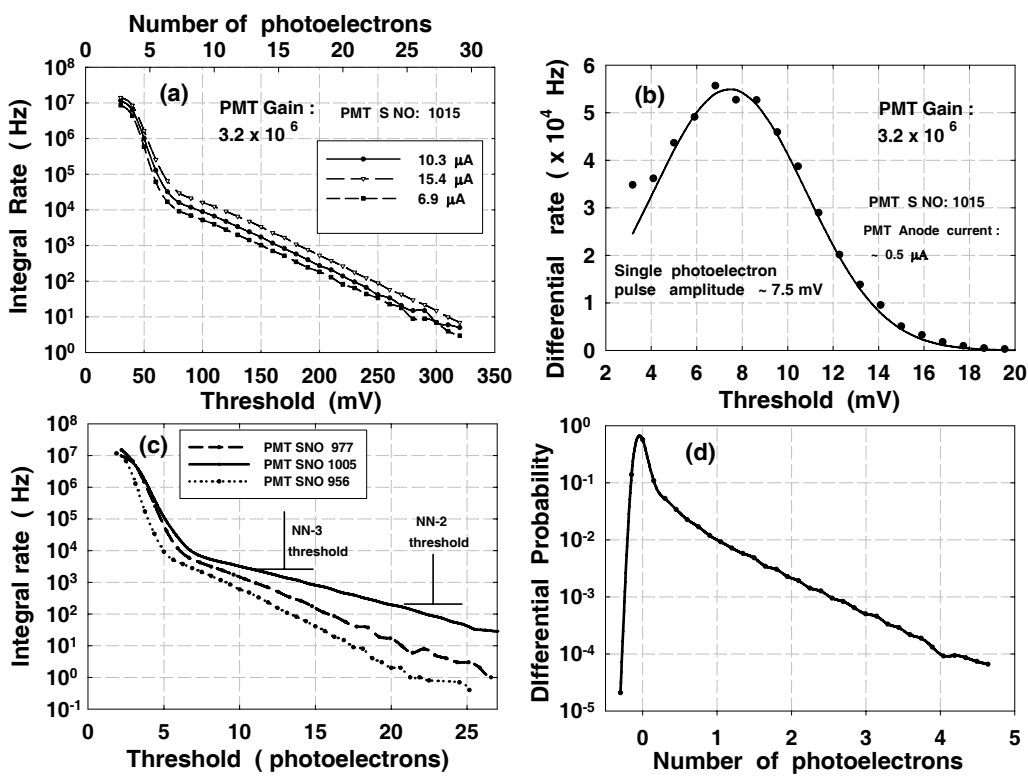}
\caption{\label{fig ---} (a)  Variation  of  trigger  rate  as a function of   discriminator  threshold  
for  3  representative  values  of light of night  sky background values.(b) Differential  count  as a 
function  of the  discriminator  threshold   for   determining  the  single  photoelectron   spectrum. (c)  
Variation  of  trigger  rate  as a function of   discriminator  threshold  for  3 different  gain 
calibrated  PMTs. (d) Measured  amplitude  distribution  of  shot-noise  fluctuations  in the output  of 
a PMT  exposed  to  a  steady  light source (like the light of night sky).}
\end{figure}
In order  to simulate   the  behaviour  of   integral  rate  when  the telescope is  observing  a 
relatively  brighter  or fainter region in the sky,  we also  show 2 more  representative curves ( marked 
as  15.4 $\mu$A and 6.7 $\mu$A) in this  figure .  The  corresponding   LONS background  values  for these 
two  curves  has been  chosen to be  $\sim 3.0\times 10^{12}$  photons $m^{-2}s^{-1}sr^{-1}$  and 
 $\sim 1.3\times 10^{12}$ photons $m^{-2}s^{-1}sr^{-1}$, respectively.   It  is  important  to mention  
here  that,   during  actual  observations,   we  operate  the  PMTs  at  a  lower  gain   and   the   
typical  values  of  anode  currents  are   $\leq$ 3 $\mu$A  for  most of  PMTs.  Operation of   PMT  at  
higher  gain  is   only  followed   during  laboratory  measurements    so that 
the PMT  can be calibrated   easily.   
The  calibration  of the   discriminator  threshold   in terms   of    photoelectrons   (top  X-axis  
scale of Fig.3a )   has   been done   by  determining   the gain  of the PMT  using  standard  single 
photoelectron detection  method.  The   measurement  for this purpose   was  taken   under  very low  
illumination level  and the  differential count rate  as a function  of  discriminator  threshold   is  
shown  in Fig.3b.    The integral  count  rate  measurements  for  3 additional   gain calibrated  PMTs   
is  shown in    Fig.3c  and  it  is   evident  from this  figure   that   the afterpulsing  rate   can 
vary  significantly   for the same  type  of PMT.  Although  a variation  of only   $\sim$$\pm$ 25\%  was  
reported  by us  in our earlier  work  [14]  on the  basis  of  about 5 PMTs  from the same  batch, 
detailed   measurements   performed,  over last several years  indicate  the  variation  in  the 
afterpulsing  rate, especially  for  PMTs  which are from different  batches,  is significantly   larger  
than  this   value.  Factors  like  variation  in the vacuum quality,  purity of the  dynode material 
used  and  the differences  in the high 
voltage  needed  to  reach the same  gain   can   be  some of the reasons  responsible  for  this  
variation.  The  differences  in the afterpulsing  rate  of PMTs   also depends on the   threshold
level  where  they  are compared.  For a large sample  of  PMTs, procured  in  different  batches, the  differences
in the  afterpulsing  rate   can be even  by a factor of  10   in  the  threshold level range of  $\sim$10-20 pes.    
Taking   guidance  from  [21],   a  function  of the type   $R_{ap}$ $\sim$ $R_{pe}\times$a $exp(-q_0 b)$ 
( where  $R_{ap}$  is the afterpulsing rate ; a  and  b are  fit  parameters)  has  also  been  fit  to 
the afterpulsing  rate  of the  3  PMTs  ( Fig.3c)  for $(q_0)>$ 5 pe.  The   best fit  values  of  b  
obtained  are  $\sim$ 0.29, 0.49 and  0.56.   A  typical  value  of  b$\sim$0.5   has  also been   
reported  for the  same  PMT type  by the HEGRA  group [21].  
\par
Referring  back  to Fig.3c,  one  can  readily   use  this  plot  to  determine  the   discriminator  
threshold    for   a predetermined  value  of  PMT  trigger rate.  For example,  taking  the  case  of  
NN-2 trigger   with 
 $C_{NN-m,n}=420$ (i.e. 121  pixels  in the trigger region),  $\tau_{c}$ $\sim$18ns  and  an additional   
condition   that 
  $f_{NN-m,n}\leq$ 0.04 Hz ( $\sim 2\%$ of  expected  cosmic ray rate  of  $\sim$2Hz );  by  using  
equation 2  we  find   that   $R_{pixel}(q_0)$ should  be   in the range  of  50  Hz to  100  Hz  with  
corresponding   values  of  $q_0$ $\sim$22pe.   On repeating   the above  exercise   for  NN-3 trigger   
  with    $C_{NN-m,n}=1840$   it  is  seen  that    $R_{pixel}(q_0)$   should  be  around  3000 Hz   with 
a corresponding  threshold   $q_0$ $\sim$ 13pe .    Keeping   in   view   that  several    key   telescope   
parameters  (e.g.   effective  collection area, threshold  energy,  differential  and integral rates)   
depend  on   the  value  $q_0$,   the above information about $q_0$  provides  a very  important  input  
for  carrying   out  Monte Carlo  simulation studies  in a realistic manner. 
 The    values  of  $q_0$,  for   the TACTIC telescope  with    NN-2  and  NN-3   trigger configurations   
have also  been marked
 in  Fig.3c .   Although  one  can determine   the  value  of  $q_0$  by performing  Monte Carlo 
simulations  at several 
 trial  $q_0$  values and  eventually  choose   the  one  for   which  the    integral rate  estimates  match 
 the  experimental  results,  following   this   approach  requires  intensive computational  resources. 
 While we  have  also followed  this  approach  but   the  main  aim  of the  exercise  was  only  for the  
purposes of fine  tuning 
 the  value  of  $q_0$   and  best  results were obtained  for the following   trigger threshold  values : 
$q_0$ $\sim$22pe    
 for   NN-2  trigger  and $q_0$ $\sim$14pe  for   NN-3  trigger.  We  will  thus   only use  these  
values  of    $q_0$ 
 for  rest of  the work  described  in the  paper. 
 \par
In order to carry out realistic simulations  for the  TACTIC telescope, 
the simulated  data-bases of the photoelectron content of each pixel in a  Cherenkov  image   also  
needs to  include the shot-noise contribution from night sky background light. Fig.3d  shows the measured  
differential  pulse amplitude  distribution  of  the shot-noise  fluctuations in the output  of a PMT.  
The  gate  pulse  to the   charge to digital converter  unit  for this measurement   was  generated   by 
using a pulse  generator.  
Using  an  appropriate   probability  density  function  for  this  noise profile  curve,  we  have  
then superimposed  the effect  of   shot-noise on the  Cherenkov  light generated  photoelectron  content  
of  each  pixel in the camera.         
\section{  Generation  of Monte Carlo simulated  data base for TACTIC  telescope  using CORSIKA code }
\label{}
For the present studies, simulations were carried out using the CORSIKA air-shower 
code (Version 5.6211) [22] 
with Cherenkov option. The CORSIKA simulation code simulates interactions of nuclei,
hadrons, muons, electrons, photons as well as   decays of unstable secondaries in the 
atmosphere. It utilises the EGS4 code for  the  electromagnetic  component of the  air shower 
simulation  and  dual parton  model  for  the simulation  of hadronic  interactions. The simulations 
have been 
carried out for Mt.Abu observatory altitude of $\sim$1300 m with appropriate values of 
35.86 $\mu$T and 26.6 $\mu$T  respectively  for the horizontal and the 
vertical components of the terrestial magnetic field.
\begin{center}
\begin{table}
\caption{ The number of gamma and proton progenitor  showers used in simulation for different 
values of primary energies at various zenith angles.}
\begin{tabular}{|c|c|c|c|c|c|}
\hline
\hline
Particle Type &\multicolumn{5}{c|}{Number of showers at various zenith angles} \\
\cline{2-6}
Gamma / Proton & & & & & \\
\hline
Energy (TeV) & $5^o$ & $15^o$ & $25^o$ & $35^o$ & $45^o$ \\
\hline
\hline

0.2 / 0.4 & 1000 / 1000 & 1000 / 1000 & 0 / 0  & 0 / 0  & 0 / 0 \\
           
\hline
0.3 / 0.6  & 1000 / 900 &  1000 / 900  & 1000 / 1000 & 0 / 0 &  0 / 0 \\

\hline
0.4 / 0.8 & 1000 / 800  & 1000 / 800  & 1000 /1000 & 0 /0 & 0 / 0 \\
      
\hline
0.55 / 1.1 & 1000 / 750  & 1000 / 700 & 1000 / 1000 & 1000 / 1000 & 0 / 0\\
       
\hline
0.7 / 1.4 & 900 / 700 & 900 / 600 & 900 /900 &900 / 900& 900 / 1000 \\

\hline
1.0 / 2.0   & 800 / 650 & 800 / 550 & 800 / 800 & 800 / 800 &800 / 900 \\

\hline
1.3 / 2.6 & 660 / 600 & 700 / 500 & 700 / 700 & 700 / 700& 700 / 800 \\

\hline
1.8 / 3.6 & 400 / 550 & 400 / 450 & 500 / 600 & 600 / 600 & 600 / 700 \\

\hline 
2.5 / 5.0 & 325 / 500 & 350 / 400& 400 / 500 & 500 / 500 & 500 / 600 \\

\hline       
3.3 / 6.6 & 300 / 450 & 325 / 350 & 350 / 450 & 400 / 450 & 400 / 550 \\

\hline
4.5 / 9.0 & 275 / 400 & 300 / 300 & 325 / 425 & 350 / 425 & 350 / 500 \\

\hline
6.0 / 12.0 & 250 / 350 & 275 / 250 & 300 / 400 & 325 / 400 & 325 / 450 \\

\hline
8.0 / 16.0 & 225 / 300 & 250 / 200& 275 / 375 & 300 / 375 & 300 / 400 \\

\hline
11.0 / 22.0 & 200 / 275 & 225 / 175 & 250 / 350 & 275 /350 & 275 / 350 \\

\hline
15.0 / 30.0 & 175 / 250 & 200 / 150 & 225 / 325& 250 /325 & 250 / 300 \\

\hline
20.0 / 40.0 & 150 / 225 & 175 / 125 & 200 / 300 & 225 / 300 & 225 / 250 \\

\hline
\hline
Total     & 8660 / 8700 & 8900 / 7450  & 8225 / 9125 & 6625 / 7125 & 5625 / 6800 \\
\hline
\hline
\end{tabular}
\end{table}
\end{center}
In order to keep the data-base  size and   computation time  within manageable limits,  
we  have  performed  multiple sampling  of  a single shower at  several core  distances. A total of 104
detector  locations  (matrix of 52$\times$2 elements), with  each detector having a size of  4m 
$\times$ 4m,  has  been used  so that  a  TACTIC-like  telescope  can sample Cherenkov light at the core 
distances of upto 245 m along 
one cartesian axis.  The  position  of the  shower  core  has  been  fixed  at  the  geometrical center  
of   the  52$\times$2 array. The  centre-to-centre inter-element spacing is  10m along the x-direction 
and 6 m along the y-direction. We have also taken care of the 
zenith angle dependence on the size of each detector element while performing the simulation at  various 
zenith angles.
Again, in order to keep the data-base size within manageable limits, the Cherenkov photon wavelength band 
chosen for 
generating the database is 300-450 nm, and the Cherenkov photon bunch  size used is varied from 1 to 30 
(depending on the energy and arrival direction of the primary particle). The simulated data-base for 
$\gamma$-ray showers uses  about $\sim$ 38035 showers in the energy range 0.2-20 TeV  with an impact 
parameter  up to 250m. These showers have been generated at 5 different zenith angles 
( i.e. $\theta$= 5$^\circ$, 15$^\circ$, 25$^\circ$, 35$^\circ$ and  45$^\circ$).   Since  it is  known  
that  the  contribution  of  cosmic  
ray  protons   is    highest   to  the   total   event rate  of atmospheric  Cherenkov   telescopes [23]  
with  other  nuclei, including   He,   contributing  up to  $\sim$30$\%$  of  the    total   event rate,  
we  have  used  only  proton primaries  in  our  simulation  study.   Taking  guidance  from  the  work 
[24], where  a similar  approach   has  been followed  in  the  simulations,    we have  scaled up   the  
event rate  calculated  for  protons  by a factor  of $\sim$ 1.5  in order  to  get  a total  cosmic   
ray  rate. Accordingly,   a data-base for  proton showers uses   about $\sim$ 39200 showers   in the 
energy range 0.4-40 TeV  and  the  incidence  angle  of these  showers  is simulated  by  randomizing  
the  arrival  direction  of the primary  in a field of view of 6.6$^\circ$ $\times$ 6.6 $^\circ$  around  the 
pointing direction of the telescope. The number of $\gamma$-ray and proton showers simulated for different 
primary energies and  at various zenith angles
are listed in Table 2. It  is  important  to mention  here  that the  number  of  gamma-ray showers   
as  well  as the  number  of  proton  showers have  not been  generated  according  to a   power law  
distribution.  The  main reason  for  this  is  to avoid   
the problem  of limited  statistics  at  higher  energies  on  account of  too  few showers. However, 
appropriate  $\gamma$-ray and  proton spectra  have  been   folded  in  the  results   while   determining  
differential  and integral  rates. Similarly, the total number  of  events(or images) considered for the 
analysis work is sampled  in an off-line manner  as per $\sim$ r dr  scaling, 
where r is the distance of the telescope from the center of the array. The total  number  of  images  
considered  for the
present  study  is  $\sim$ 1.9 million   for  $\gamma$-rays  and   $\sim$ 2.0 million  for   protons.    
\par
The  CORSIKA  code  generates   a  Cherenkov file  containing   seven  parameters  for   characterizing  
the Cherenkov photon (or photon bunch)  received  with  a detector  element  area (4m$\times$4m  in our 
case).  These parameters are bunch number, x and y 
co-ordinates, two direction cosines, height of production and arrival time at the  ground. This CORSIKA 
code which we have used does not consider wavelength dependent atmospheric absorption of  the Cherenkov 
photon (bunch). Accordingly, atmospheric absorption of the Cherenkov  photon-bunch as well as its 
subsequant history, is taken care of by a supplementary 
BACUP-code, developed in house. Essential details related of this code are described  below. 
\par
The parameters of each photon-bunch, produced in the wavelength range, $\lambda \sim$ 300-450 nm by 
simulation code CORSIKA and managing to the reach 52$\times$2 detector array, are written in the 
Cherenkov file. The BACUP-code uses 
this information to obtain corresponding parameters of photons expected in the 
overall wavelength region $\lambda \sim$ 200-600 nm. Keeping in view the variations 
in the atmospheric extinction coefficient at shorter and longer wavelength, the 
step-size used is 5 nm, 10 nm, 20 nm and 50nm for the ranges $\lambda$ = 200-270 nm, 
270-280 nm, 280-400 nm and $\lambda \ge$ 400 nm respectively. The number of Cherenkov 
photons  expected in various wavelength intervals is extrapolated using $\sim$ 
$\lambda^{-2}$ dependence of the Cherenkov light emission spectrum.  Each  Cherenkov photon  
is traced back to the point of emission and the $\lambda$- dependent atmospheric 
extinction is considered for the  wavelength range $\lambda$ = 200-270 nm, 
and $\lambda \ge$ 270nm, using   measured   data   and  using data derived from [25] respectively. 
Each photon-bunch, after  correcting for  atmospheric absorption, is reflected from the mirror-facet and 
ray-traced to the corresponding 
position in the focal plane of the   telescope.  Appropriate care is taken to 
include the  reflectivity of  mirrors, transmission of the light guides (CPC) and  
quantum efficiency of the phototubes in the present simulation 
studies.  A modest  value  of  $\sim$82$\%$ for the  photoelectron  collection efficiency  has  
been  used  in teh simulation. 
The obstruction encountered due to the telescope mechanical structure by 
the incident and reflected photon-bunches, during their prorogation, is also duly 
considered. The data-base of  resulting photo-electrons (pe)  in the imaging camera pixels, are subjected 
to noise injection and  image cleaning  for further analysis.   Representative  examples   of   
Cherenkov  images
recorded  at a zentih  of angle 25$^\circ$  is   shown in  Fig. 4.    
\begin{figure}[h]
\includegraphics*[width=0.9\textwidth,angle=0,clip]{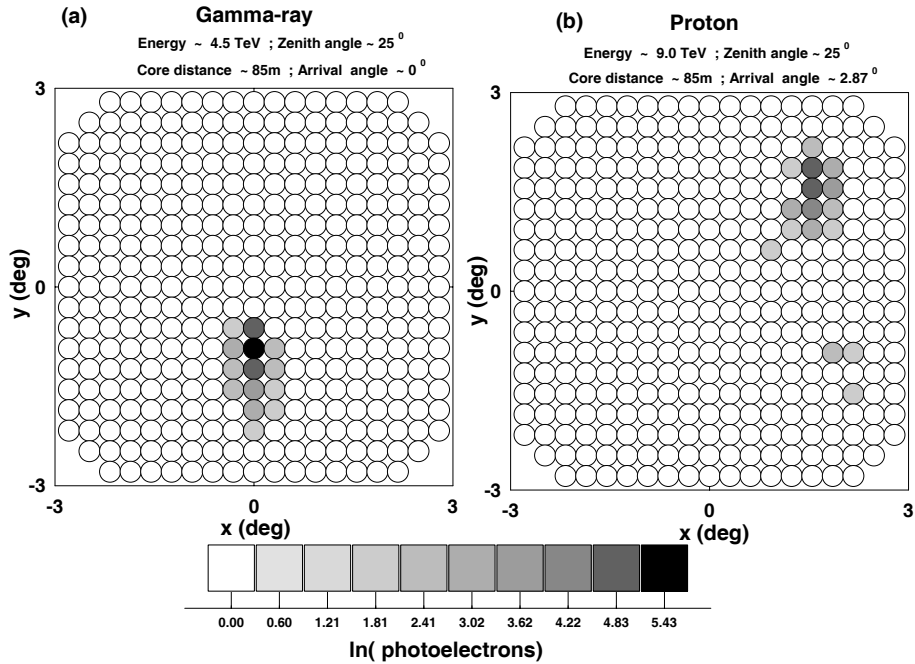}
\caption{ (a)  Cherenkov  image  of a  $\gamma$-ray  shower  with  energy $\sim$4.5  TeV and a core 
distance of $\sim$ 85m. (b) Cherenkov  image  of a  proton   shower  with  energy $\sim$9.0 TeV, core 
distance of $\sim$ 85m  and  arrival angle  $\sim$2.87$^\circ$.}
\end{figure}
The  grey  shade color  scheme  followed,  for  representing   number of photoelectrons   recorded in  a  
pixel, using  equal ln(photoelectrons) intervals, is also shown at  the bottom  of this  figure.  The  
Cherenkov  image  of a  $\gamma$-ray  shower,  shown  in Fig.4a,    has   a primary  energy of  
$\sim$4.5 TeV  and a core distance of $\sim$ 85m.  The   Cherenkov  image,  shown  in  Fig.4b, for  an 
off-axis proton  primary   has  the following  parameters : energy $\sim$9.0 TeV ,arrival  direction 
$\sim$2.87$^\circ$ and  core distance of $\sim$ 85m.
\section { Formalism  for calculating  effective  collection rate, threshold energy  and  detection rates} 
\label{}
The   trigger  rate  of an atmospheric  Cherenkov  imaging   telescope  and the resulting  statistics  
of the showers is  primarily  decided  by two  basic parameters;  effective collection area  and the 
energy  threshold.  For a given configuration
of the telescope, these parameters  depend on the  pixel trigger threshold,  the trigger multiplicity and  
the  trigger  field of view. The   energy threshold  of the telescope   primarily  depends  on the 
geometrical  area  of the  light  collector and the photon to electron  conversion efficiency.  The later  
being  determined   by the   mirror  reflectivity,  collection efficiency of the  light guides, the spectral
  response of the  PMTs  and other possible  losses  of the  Cherenkov light  on the  way to the PMTs. 
\par
The  effective  collection   area  of an atmospheric  Cherenkov  imaging  telescope  is   governed  by the 
lateral  and the  angular  distribution  of the  Cherenkov  light   generated  by the primary  particle 
(i.e. $\gamma$-ray  or  a cosmic-ray).  For a $\gamma$-ray, coming  from  a point  source,   the  effective  
collection area ( $A_{\gamma}(E)$) is   given  by
\begin{equation}
A_{\gamma}(E) =  2 \pi  \int_{0}^{\infty}    P_{\gamma}(E,r) r dr \
\end{equation}
where  $P_{\gamma}(E,r)$  is the detection  probability  (or  trigger efficiency) for a $\gamma$-ray  
 with  primary energy  E  and  
impact parameter r. By the trigger efficiency $P_{\gamma}(E,r)$, we imply the fraction of the simulated 
Cherenkov events which can be detected by the  telescope  placed at a given distance r from $\gamma$- ray 
or proton- generated EAS core.
Since, for the  isotropic  cosmic-ray background  the   detection  probability    $P_{CR}(E,r,\Theta)$ 
depends  also  on the  angle  between the direction of the primary particle  and  the telescope axis 
($\Theta$), the   effective  collection area ( $A_{CR}(E)$) is   given  by 
\begin{equation}
A_{CR}(E) =  2 \pi  \int_{0}^{ 2\pi }  \int_{0}^{ \infty}  P_{CR}(E,r,\Theta)   r dr d \Omega  \
\end{equation}
where  $d\Omega=2 \pi \sin \Theta$  and   the  integration  over  $\Theta$  is included  in  $d \Omega$.   
It is important   to mention that  the effective collection area   for  cosmic  rays,  given by  
equation 6,  has    units  of   $m^{2}$ $sr$. 
Using  the  effective  collection area  derived  from above equations, the differential  rate   can   
be  found  by   multiplying   the  effective  area  with  the   corresponding   differential  spectra  
of  $\gamma$-rays  and cosmic-rays ; 
i.e $dR_{\gamma}/dE= A_{\gamma}(E) \times  dF_{\gamma}/dE$     and   
$dR_{CR}/dE= A_{CR}(E) \times  dF_{CR}/(dE d\Omega)$.       
For  $\gamma$-rays   we have  used  the  Crab Nebula  differential  spectrum  [26]
\begin{equation}
\left(\frac{dF_{\gamma}}{dE}\right)= 6.0 \times 10^{-6} E^{-2.31-0.26\times log\left(\frac{E}{300GeV}\right)} m^{-2}s^{-1}TeV^{-1} 
\end{equation}
For   the   cosmic-ray  background,  we have used  the following  spectrum [24]      
 \begin{equation}
\left(\frac{dF_{CR}}{dE d\Omega}\right)= 0.096 E^{-2.7}m^{-2}s^{-1}sr^{-1} TeV^{-1} 
\end{equation}
As  already  mentioned in   section 5,  we  only  use  proton showers   for   cosmic-ray  simulations.  
The reason  for this being   the  fact   that   for a given energy, the showers induced  by heavier  
nuclei  develop  at  substantially   larger   heights   in the atmosphere  with a less  intense  Cherenkov  
light pool.  In addition, the fraction of energy  channeled into  electromagnetic  subshowers,  responsible  
for  the  production of Cherenkov  photons  decreases  with   increasing   nucleon 
number.   A  combination of  all  these   result  in a larger  but  less intense  Cherenkov light  pool   
which  increases  the  
threshold  energy of the  telescope   for  heavier  cosmic  ray  particles.
The  integral  trigger rate for   $\gamma$-rays ($R_{\gamma}$) and  cosmic-rays ($R_{CR}$) is   given by 
\begin{equation}
R_{\gamma} = \int_{ 0}^{  \infty} \frac{dF_{\gamma}}{{dE}}A_{\gamma}(E)\ dE \
\end{equation}
\begin{equation}
R_{CR} = \int_{0}^ \infty  \frac{dF_{CR}}{dEd\Omega}\ A_{CR}(E)\ dE \
\end{equation}
It is  important  to mention here  that   we have  discretized  the  integrals ( equations 5  and 6)  in 
our  simulations  with a step  size of  10m  in core distance    and  a step size  of  $\sim$0.1$^\circ$ for   the  arrival  direction. The  largest  value  of  core  distance  used  in the  simulation  is  $\sim$ 
245m ( imposed by using an array of 490 m length in the CORSIKA simulations) and   this  restriction yields 
reasonably accurate values for primary energy range considered in the present work. 
Significant deviations in calculation of the effective area are expected at higher 
primary energies ($>$ 40 TeV), but the relative contribution of such particles 
is only marginal.
\section {Optimization of the trigger field of view}
\label{}
While   pixel   size  and the total  field  of  view  are the basic parameters  in the design of a camera  
  for an   imaging telescope,  it  is also  important  to optimize  the  trigger  field of view  so that, 
$\gamma$-ray showers   with  an impact parameter  of up to  150m, are able  to   trigger  the  telescope.  
 Keeping   in view  that  the  detection rate of showers  initiated  by isotropically   distributed  cosmic  
rays is  proportional  to the  solid angle, using  a  trigger  field of view   beyond  the optimum value  
only  reduces  the ratio  of  $\gamma$-ray  trigger rate  to  cosmic-ray  trigger rate.  The optimum  
trigger  field  of  view  is decided
 by the position of the  image centroid  in the camera  and  is  determined  by the  height of the shower  
maximum  above   the  observatory level and  the impact parameter.  Knowing that,   for  $\gamma$-ray 
showers,  the    angular  shift  of the image  from the camera center can be  estimated  quite  reliably  
using a  simple  toy  model of the  electromagnetic   cascade development,  it is  worth   having  a look   
first  at  these  estimates   before  proceeding  to   detail Monte Carlo simulations.  In the toy  model, 
one assumes  that  most  of the Cherenkov  light  photons are coming  directly  from   the shower  maximum. 
If    $H_{max}$   is  the height of shower maximum   (with   $\chi_{max}$   as   the corresponding  
atmospheric  depth)    where  the number of  secondary  electrons  in the shower is maximum,   the  
expression  for    $\chi_{max}$,    using   classical   cascade  theory  is given  by :       
 \begin{equation}
 \chi_{max} \approx   \chi_{0}  \ln [ (E_{\gamma}/E_c)]  \\
\end{equation}
where   $E_{\gamma}$ is  the primary  energy  of the  $\gamma$-ray;   $E_c$  is   the  critical energy  
($E_c$ $\sim$ 76 MeV)   and  
$\chi _{0}$  is the radiation  length  in the atmosphere ($\chi _{0}$ $\sim$ 36.6 g$cm^{-2}$).  The   
angular  displacement  of the  image  centroid  from the camera  center   depends  on the   impact  
parameter  and   the   line  of sight  distance  to the  shower  maximum and  is  given  by the following    
expression [2]:  
 \begin{equation}
 \delta_{max} \approx    \tan^{-1}  \left[\frac { b \cos \theta } 
{ b \sin \theta  +  h_{0}  \ln [1030/( \chi_{max} \cos \theta)] - H_{obs}}\right]  \\
\end{equation}
where  $h_0$  is the   scale  height  of the atmosphere  ( $h_0$ $\sim$ 8.5 km ) ; $H_{obs}$  is   the 
observatory  altitude;  b  is the   impact  parameter   and  $\theta$    is the  zenith  angle.   The  
variation  of  angular shift   of the image  centroid  as a function  of impact parameter   for   different 
$\gamma$-ray energies  is shown in  Fig.5.   
\begin{figure}[h]
\includegraphics*[width=0.9\textwidth,angle=0,clip]{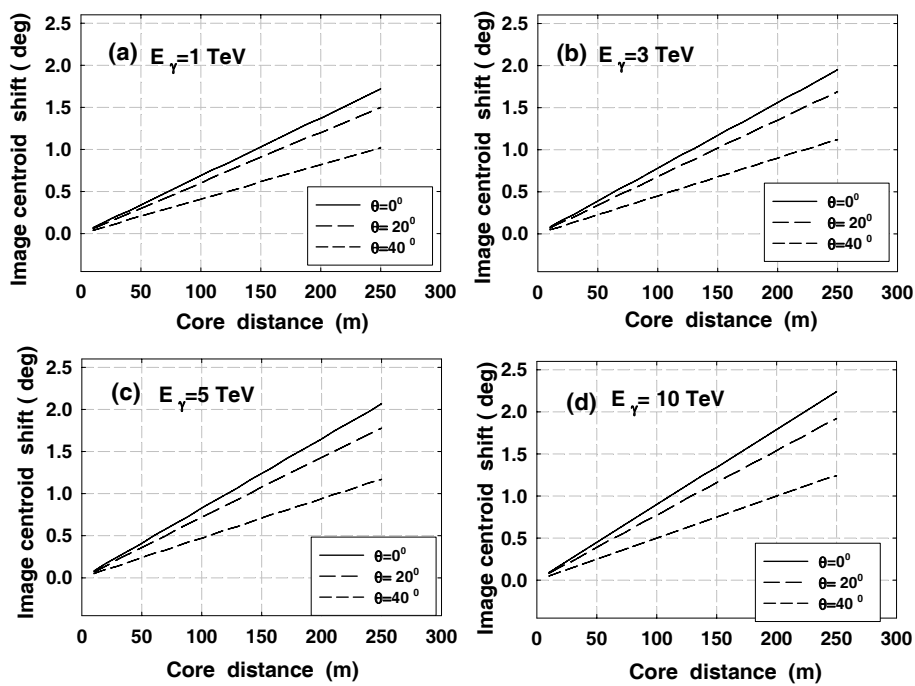}
\caption{ Angular  displacement  of the  image centroid  from the camera  center  as a function of core 
distance  for  various  zenith angles ( $\theta$=  $0^o$, $20^o$ and  $40^o$) and   primary $\gamma$-ray.  
(a)  $E_{\gamma}$=1 TeV  (b) $E_{\gamma}$=3 TeV  (c) $E_{\gamma}$=5 TeV  and  $E_{\gamma}$=10 TeV.}
\end{figure}
On examining  this  figure, it  is  clearly  seen  that   an imaging  camera  with a  trigger  field  of  
view   of  $\sim$ $1.5^o$   (half-angle) is  able  to   detect  $\gamma$-ray  showers of  energies  
$<5TeV$   up to   impact  parameter  values of   $\sim$$150m$.  It  is  also seen  in the  figure   that   
the  angular shift   of the image  centroid  from  the  camera  center  decreases  when  zenith  angle is 
increased.  Furthermore, if
the  main   concern  is to  maximize  the  ratio  of  $\gamma$-ray  trigger rate  to  cosmic ray  trigger 
rate,  further   increase  in   the  trigger field  of  view  does  not  lead  to  an  increase  in the 
$\gamma$-ray trigger rate   because of the  fact that  the   detection  of   $\gamma$-ray showers  beyond   
$\sim$ $150m$  is   limited   due  to the  reduced  density  of the  Cherenkov  photons,  
\par
For optimizing  the  trigger  field  of view  on the basis  of  Monte Carlo simulations   we  have  used  
simulated  data   generated  at two  zenith  angles (viz., $15^o$  and  $35^o$) with  trigger  regions   
having    49, 81, 121, 169  and 240  pixels.  The  corresponding   fields  of  view  of  these   trigger 
regions  are $\sim$ $2.17^o$ $\times$ $2.17^o$,   $\sim$ $2.79^o$ $\times$ $2.79^o$,   
$\sim$ $3.41^o$ $\times$ $3.41^o$,  $\sim$ $4.03^o$ $\times$ $4.03^o$  and  
$\sim$ $4.65^o$ $\times$ $4.96^o$.  Both  NN-2  as  well as NN-3   trigger  generation  schemes,  with  a    
single  pixel  threshold   ( q$_{0}$)  of  $\sim$ 14pe and   $\sim$ 22pe, respectively,   have   been   
considered   in  the  simulation  study.      The  trigger  efficiency  as  a function   of  $\gamma$-ray  
energy,  for  showers  falling   with  a core distance  of  $\sim$250m   is shown  in  Fig.6.   Referring  
first  to  Fig. 6a and Fig. 6b ( i.e  efficiency  curves  at a zenith angle  of  $15^o$  for  NN-2 and  
NN-3  trigger modes), it  is evident  that  at  $E_{\gamma}$ $\geq$ 4.0 TeV,  the   trigger  efficiency    
increases   when  the  number of pixels  in the   trigger  region  is  increased   from  49  to 121.  
It is also seen  that  the   trigger  efficiency remains  almost  constant    when   the   number  of  
trigger  pixels  is  increased  further  from  121 to 240.   A similar  behaviour   is   also  evident  in   
Fig.6c  and  Fig.6d   when   a  zenith  angle of  $35^o$  is  used  for  studying   the  trigger  
efficiency.       
\begin{figure}[h]
\includegraphics*[width=0.9\textwidth,angle=0,clip]{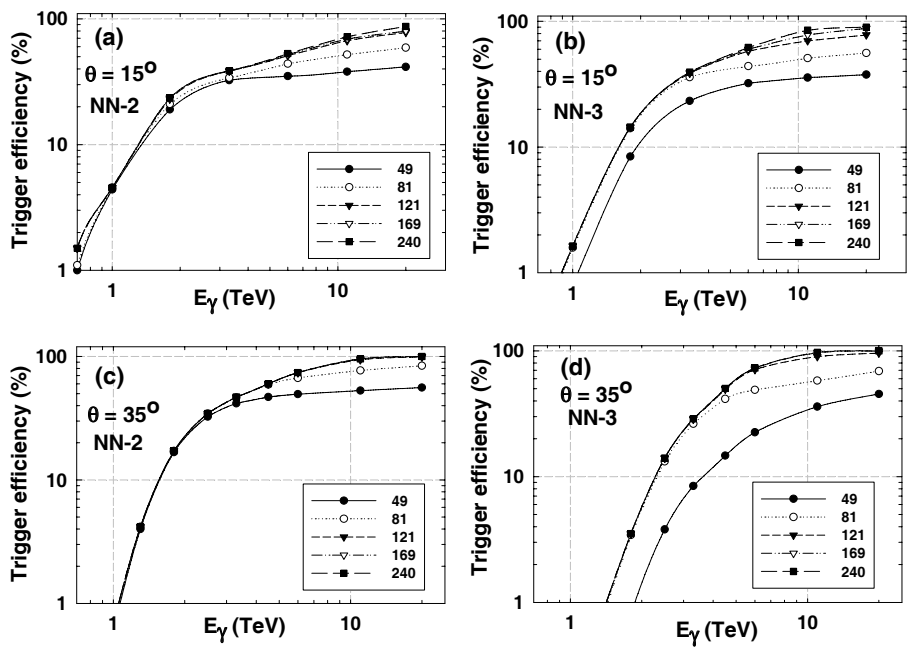}
\caption{Trigger efficiency  at  a  function   of   energy of the  primary  $\gamma$-rays  with  trigger  
regions   having    49, 81, 121, 169  and 240  pixels. (a)  NN-2  trigger  with   $\theta$=15$^o$,     
(b)  NN-3  trigger  with   $\theta$=15$^o$ (c)  NN-2  trigger  with   $\theta$=35$^o$  and    (d)  NN-3  
trigger  with   $\theta$=35$^o$. }
\end{figure}
Therefore,  these  results  suggest  that  using   121  pixels  ( 11$\times$11 matrix), having a field  of 
view  of    $\sim$ $3.41^o$ $\times$ $3.41^o$,  is  quite optimum  for  producing   triggers   in response   
to  $\gamma$-rays  with  an  impact  parameter   up to $\sim$ 150m.  
\par
We   have   also   estimated   the  rate  loss   for  cosmic rays   when   the   largest  value  of  core distance  used in
the  simulations  is  restricted  to  $\sim$245m.  This  restriction  arises  because  of  using  an  array of  490m  length in the CORSIKA  simulations.  In  this  study   we have  included   showers  with  an impact parameter   up to  300m  and   the  behaviour  of  the  cosmic ray  rate  as a function  of  core distance  is  shown  in  Fig.7(a-d).  Referring  
first  to  Fig. 7a and Fig. 7b ( i.e  cosmic ray  rate  at a zenith angle  of  $15^o$  for  NN-2 and  
NN-3  trigger modes),  one  can   easily  see   that,   for  a  trigger  field  of  view with  240  pixels,  there   is  only a  marginal   increase  in  the  cosmic  ray rate   ( $\sim$2.5$\%$)    when  the  core  distance    increases  from 
$\sim$245m   to  $\sim$300m.  It  is  also  seen  in  the  figure   that  corresponding   increase  in   the   cosmic ray rate  is   negligible   for  a  trigger  field  of  view of  with  49  pixels.  A  similar  behaviour   is   also  seen
in   Fig.7c  and  Fig.7d   when   a  zenith  angle of  $35^o$  is  used.          
\begin{figure}[h]
\includegraphics*[width=0.9\textwidth,angle=0,clip]{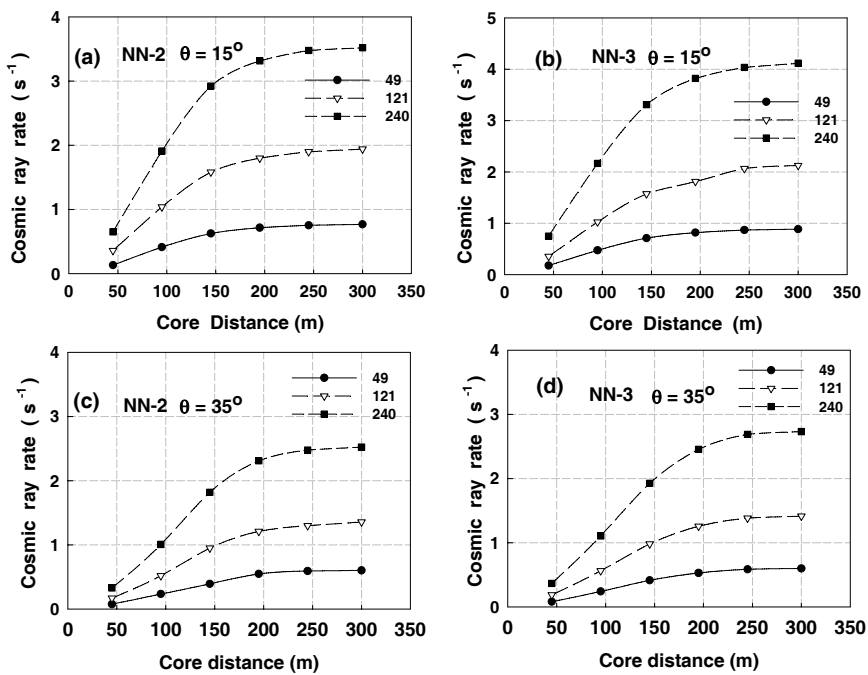}
\caption{Cosmic  ray  rate  as   a  function   of   core  distance  with  trigger  
region   having    49,  121  and 240  pixels. (a)  NN-2  trigger  with   $\theta$=15$^o$,     
(b)  NN-3  trigger  with   $\theta$=15$^o$ (c)  NN-2  trigger  with   $\theta$=35$^o$  and    (d)  NN-3  
trigger  with   $\theta$=35$^o$. }
\end{figure}
\par
The   performance  of  the  imaging  camera  with   49, 81, 121, 169  and 240  pixels  in the  trigger 
region  has  also   been  studied   by  calculating   the  integral  $\gamma$-ray  
and  cosmic-ray  trigger rate   and   the  results  of  this study  are  shown in Fig.8. 
\begin{figure}[h]
\includegraphics*[width=0.9\textwidth,angle=0,clip]{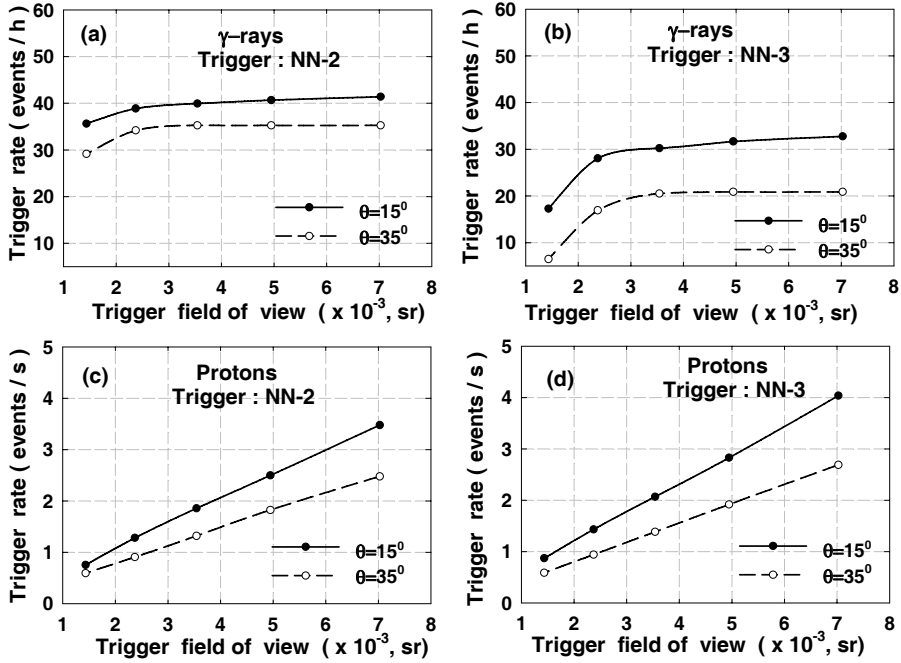}
\caption{ (a $\&$ b) Gamma-ray   trigger  rate   as  a function  of  trigger field of  view for  NN-2   and  
NN-3 trigger schemes. 
(c $\&$ d) Cosmic-ray   trigger  rate   as a function  of  trigger field of  view  for  NN-2  and  NN-3 
trigger schemes. }
\end{figure}
The  $\gamma$-ray  trigger  rates   shown in Fig.8a  and Fig.8b  have  been  calculated  by using   
the Crab nebula energy  
spectrum. The  corresponding  trigger rates    for  cosmic-ray  protons  as a function  of   trigger  
field  of view  are  shown in  Fig.8c  and Fig.8d. Strong    dependence  of  the $\gamma$-ray  trigger  
rate  as  a function of  trigger  field  of view   is  only seen  when  49  pixels  are  used   for  
trigger  generation.  A   weak  dependence   of   $\gamma$-ray  trigger  rate,   for  all  other 
configurations with  $>$ 121  pixels in the trigger region,  is  also  evident  from these  figures.  With   
trigger  rates of  
$\sim$ 40 events h$^{-1}$  for  NN-2 trigger    and  $\sim$ 30 events h$^{-1}$  for  NN-3 trigger   at a 
zenith  angle of 15$^o$,  it  is  also  seen  from these  figures  that   NN-2  trigger  scheme  provides   
better  detection efficiency   at  higher zenith angles  as compared  to the  NN-3 trigger  scheme.   
Linear  dependence   of the  proton  trigger  rate  on  the trigger  field of view   seen in  Fig.8c  and 
Fig.8d,  in accordance  with  the   isotropic  behaviour   of  cosmic  rays,  also   reassures  that  the  
trigger  field  optimization  results   are reliable.  Therefore it can  be  safely concluded   that   a 
trigger region  using   innermost  121   pixels  in the TACTIC  imaging camera  provides an optimal  field  
of view  for  detecting $\gamma$-rays   from  a point  source.   
\par
It is important  to mention   here  that the  actual  field of view  of the camera   should  be more  than  
the  trigger field so that  truncation  of  the  Cherenkov   images  can  be avoided [21].  The  shapes  
of   Cherenkov  images,  produced by  $\gamma$-rays,   get otherwise  distorted   and  the effect  is  
especially   important  for the LENGTH  parameter  (defined as the rms   angular size  and is related  to 
the longitudinal  development  of the  shower).  The  limited  field  of view  also   introduces  large  
errors  in the determination  of  other  image parameters  and   in  estimating   the  energy  of the 
primary  $\gamma$-rays.  It  is  for this  reason   that  imaging   camera  of the  TACTIC telescope   
uses   225 to  349 pixels.   Further   work  related  to  realizing  the  potential  of the full   camera  
with  349 pixels,  especially  using  Wobble  mode  of   observation   is   still in progress.
\section{Optimization of topological trigger generation  scheme  on the  basis  of  estimated threshold  
energy  and  detection rates}
\label{}
After    having   validated     that   a trigger field  of  view 
with  121  pixel  is   quite  optimum  for detecting the 
$\gamma$-rays  from a point  source, we now proceed to   performance evaluation of the TACTIC   telescope  
on the  basis  of  detection rates  of   $\gamma$-rays  and  cosmic-rays.  Optimization  of  the topological 
trigger condition by  considering
2 and  3  fold  nearest neighbour   trigger schemes  (i.e.  NN-2  versus NN-3) is   another  aspect which  
is   discussed  in  this  section.  Collection  areas  for  $\gamma$-ray   and  
cosmic-ray  induced  showers   can be  derived  from  Monte Carlo  simulations    
by  evaluating   the  energy  dependent  trigger efficiency ($P_{\gamma}(E,r)$ and $P_{CR}(E,r,\Theta)$).   
Using the equations 5 and 6, along with  a   single pixel  threshold  of  $\sim$ 22pe for the  NN-2  trigger  
and   $\sim$ 14pe for the  NN-3,  the Monte Carlo evaluated effective collection areas,  as a function of 
the  progenitor particle-energy, are shown in Fig.9 for 3-different zenith angles. The results  have  
been   obtained   by    using  the inner-most 121 pixels (11$\times$11 matrix) for trigger   generation. 
\begin{figure}[h]
\includegraphics*[width=0.9\textwidth,angle=0,clip]{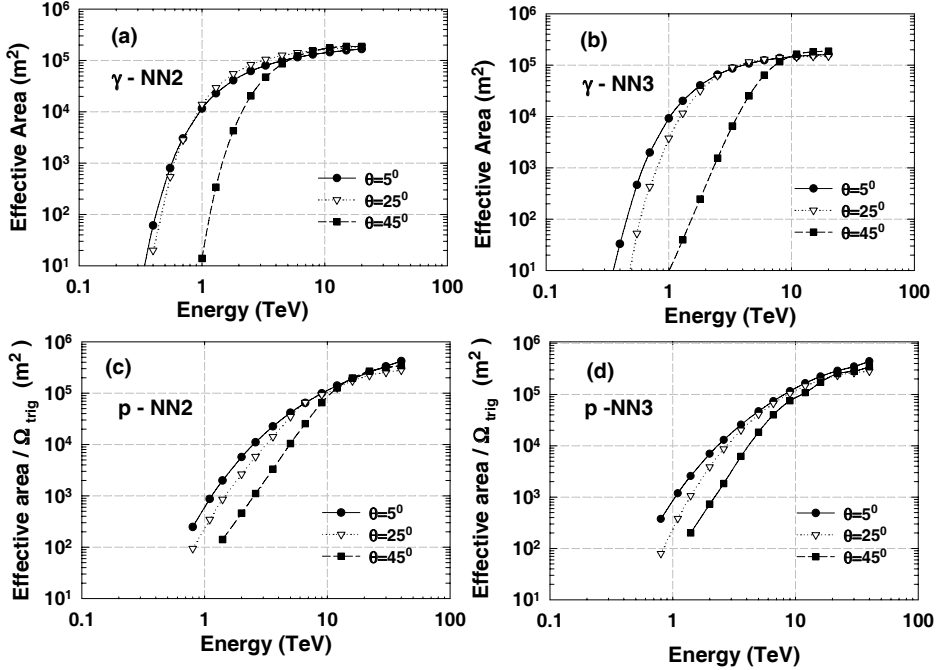}
\caption{ Effective collection areas for $\gamma$-ray and proton induced showers, 
at various incident angles, with NN-2 and NN-3 trigger schemes. (a) $\gamma$-rays   with  NN-2  trigger 
scheme (b)$\gamma$-rays   with  NN-3  trigger scheme.  (c) Protons   with  NN-2  trigger scheme.(d) 
Protons with  NN-3  trigger scheme.}
\end{figure}
To facilitate   an easier   comparison of  effective  areas, we  have  divided  the  effective collection 
area for protons  by    $\Omega_{trig}$, where $\Omega_{trig}$ 
denotes the trigger  field of view  solid angle ($\sim3.54 \times 10^{-3}sr$).  
The energy  dependence  of the  effective  collection
area   on the  primary  energy  is  clearly evident   in this  figure  and  reflects  the  shape  of the  
lateral  distribution  of  Chrenkov photons on the ground. Compared with  the relatively  steep lateral 
distribution  of the  cosmic-ray  showers, the  lateral  distribution  of   Cherenkov  light  produced  
by $\gamma$-ray  showers   has  a  rather  flat lateral  distribution   up to  $\sim$120m   core distance  
( a more  precise value  is  dependent  on the observatory  altitude).   
Using  derived effective collection areas shown  in Fig.9, the  corresponding   differential  rates,
for   $\gamma$-rays  and  protons  is  shown in Fig.10(a-d).
\begin{figure}[h]
\includegraphics*[width=0.9\textwidth,angle=0,clip]{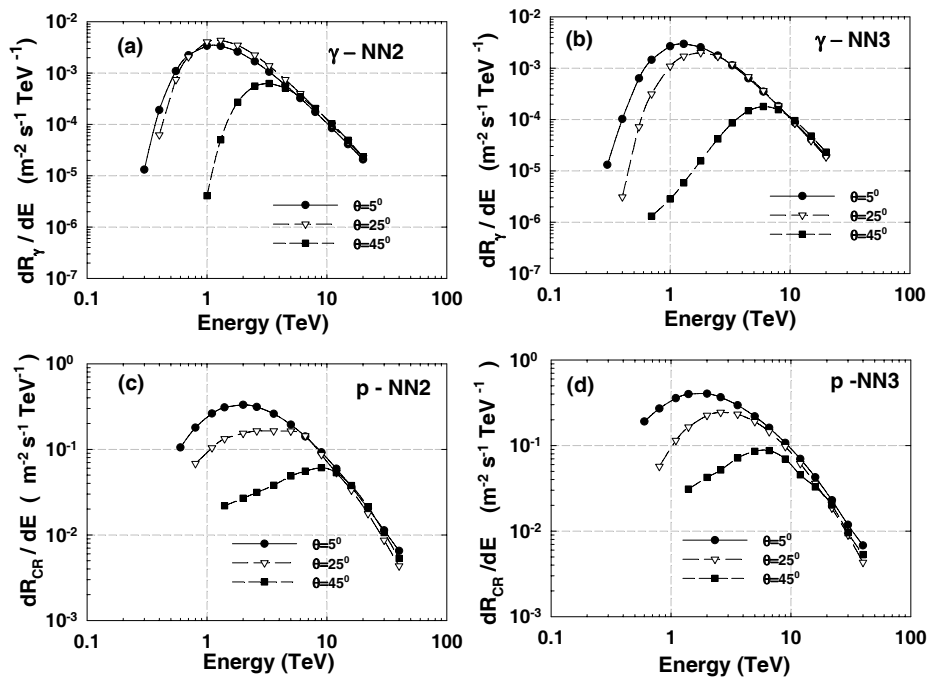}
\caption{ Estimated differential rates  for   $\gamma$-ray and proton induced showers, 
at various incident angles, with NN-2 and NN-3 trigger schemes ; (a) $\gamma$-rays   with  NN-2  trigger 
scheme  (b)$\gamma$-rays   with  NN-3  trigger scheme; (c) Protons   with  NN-2  trigger scheme ;(d) 
Protons with  NN-3  trigger scheme.}
\end{figure}
The  differential  rate  curves, for the   two event  species,  have  been  obtained  by  using   Crab 
Nebula spectrum for the $\gamma$-rays  (equation 7 ) and  proton spectrum ( equation 8). 
Following   the standard  definition  of   effective  threshold 
of the telescope i.e.,  value of the primary  energy at which    the  differential    rate  peaks,  the 
resulting  energy  thresholds of   the  TACTIC   telescope    are  presented  in  Table 3.    
Results   regarding  integral  rates, calculated  by using  equations 9 and 10, are   shown  in Fig.11  
and  also  given  in    Table 3.  It  is important  to mention  here that we have  scaled up  the  calculated  event 
rate  for  protons  by  a   factor  of $\sim$ 1.5  in   order  to  estimate  the   total  
cosmic   ray  rate ( as  discussed  in section 5).  The  main  purpose  of  this  scaling  is  to  account   for  contributions  from   other  nuclei, including  He.    
\begin{figure}[h]
\includegraphics*[width=0.9\textwidth,angle=0,clip]{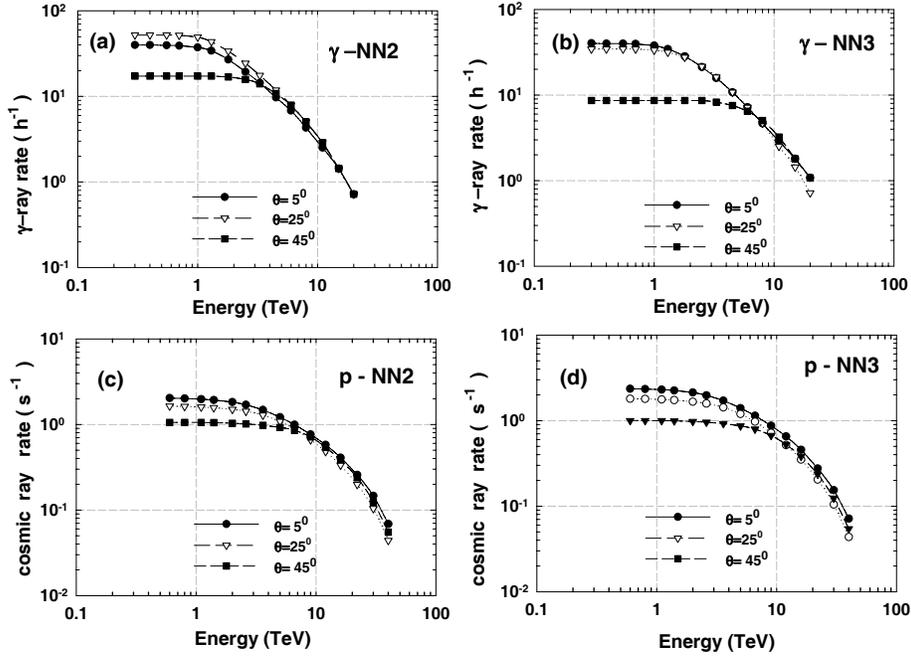}
\caption{  Estimated  integral  rates for   $\gamma$-ray and cosmic ray  induced showers, 
at various incident angles, with NN-2 and NN-3 trigger schemes. (a) $\gamma$-rays   with  NN-2  trigger 
scheme  (b)$\gamma$-rays   with  NN-3  trigger scheme.  (c) Cosmic-rays   with  NN-2  trigger scheme.(d) 
Cosmic-rays with  NN-3  trigger scheme.}
\end{figure}
\begin{table}
\centering
\caption{  Threshold energy  for
$\gamma$-rays (Crab-nebula spectrum) and  protons  for the TACTIC  telescope, with  121  pixels in the  
trigger  region  and  using NN-2 and NN-3 trigger configurations. The integral detection rates  for  
$\gamma$-rays and cosmic-rays (CR) are  also given in the table. 
 A scaling  factor  of $\sim$ 1.5  ( as  discussed  in section 5)  has  also been  used  to  estimate  
the  total  cosmic-ray  rate.}
\begin{tabular}{|c|c|c|c|c|c|c|c|c|}
\hline
\hline
Zenith angle   & \multicolumn{4}{c|}{ Energy Threshold (TeV)}  &  \multicolumn{4}{c|}{Integral Rate } \\
\cline{2-9}
(in degree) &\multicolumn{2}{c|}{ NN-2} &\multicolumn{2}{c|}{ NN-3}&\multicolumn{2}{c|}{ NN-2} 
&\multicolumn{2}{c|}{NN-3} \\
\cline{2-9}
& $\gamma$ & Proton &  $\gamma$ & Proton & $\gamma$( h$^{-1}$)  & CR ( s$^{-1}$)& $\gamma$ (h$^{-1}$)  & 
CR (s $^{-1}$)  \\

\hline
\hline
5 & 1.3 & 2.6 & 1.3 & 2.6 & 40 & 2.0 & 40 & 2.4 \\
\hline
15 & 1.3 & 3.6 & 1.8 & 3.6 & 36 & 1.9 & 34 & 2.1 \\
\hline
25 & 1.3 & 3.6 & 1.8 & 3.6 & 50 & 1.6 & 34 & 1.8 \\
\hline
35 & 1.8 & 3.6 & 2.5 & 3.6 & 36 & 1.3 & 22 & 1.4 \\
\hline
45 & 3.3 & 9.0 & 6.0 & 9.0 & 18 & 1.0 & 11 & 1.0 \\
\hline
\end{tabular}
\end{table}
\par
On comparing   the  performance  of  NN-2 and NN-3 trigger  configurations,  one  can  notice  the  
following :   (i) In  the  case of  NN-3  trigger,  there is a marginal increase in the $\gamma$-ray 
threshold  energy  when  zenith  angle  is increased  from  5$^o$ to 25$^o$. (ii)   Beyond 
$\theta$$>$25$^o$  the  $\gamma$-ray  threshold energy increases more rapidly for NN-3 trigger 
($\sim$6.0TeV at 45$^o$)  as compared  to  the  NN-2  trigger. In 
the  case  of NN-2 trigger  the  $\gamma$-ray  threshold  energy  changes  from 1.8TeV  at 
$\theta$=35$^o$ to   $\sim$3.3TeV  at   $\theta$=45$^o$.  The  resulting  integral trigger rate   for  
$\gamma$-rays as a function of  zenith  angle  is also   seen   to  drop more   rapidly in NN-3  trigger 
mode  compared  to  the   NN-2  trigger   configuration. The   reason  for   this  being  the fact  that  
at larger  zenith angles  the Cherenkov    images produced  by  $\gamma$-rays   tend  to be more  compact    
as  a result  of  increase  in the  line  of sight  distance  to the  shower  maximum. 
Using   a  NN-2  trigger  methodology  is   thus  a  better option  as  against   the   NN-3  trigger  
configuration  when  a   source    needs to be observed  at   $\theta$$>$ 30$^o$.   Advantage  of   using  a  
NN-2 trigger  configuration,
for the   HEGRA  telescope   with  a pixel size  of   $\sim$ 0.25$^o$,
has  also been  pointed  out  in [21]  while  comparing  the performance of various   nearest
neighbour  trigger  schemes  with  different  multiplicities. 
\section{Comparison with experimental data}
\label{}
The 349-pixel TACTIC imaging telescope, 
has been  in operation at  Mt.Abu, India  since 2001  and   has  so far 
detected $\gamma$-ray emission from  the Crab Nebula, Mrk 421 and Mrk 501. First  successful detection  of  
TeV $\gamma$-rays from  the Crab Nebula was  seen  in the  data  collected by the   telescope  for 
$\sim$ 41.5h  between January 19 - February  23, 2001.
The data  was  taken  by using  a  NN-3  topological trigger 
with the innermost 240 pixels (15 x 16 matrix) of the camera  participating in generating the  event-trigger.  
A  statistically significant excess of $\sim$6.3$\sigma$  was  seen  in  the  data  
with  $\sim$447$\pm$71  $\gamma$-ray like events.   During the last few  years,  regular observations were  
taken on a number of potential $\gamma$-ray sources  (viz., 1ES2344+514, PSR 0355+54, ON 231, H1426, 
Mrk-421, Mrk-501 etc.)    and   the results of these  observations  have  been  presented in [27-30]. A  
major  portion  of   this data   has been collected  with  121 pixels (11$\times$ 11 matrix) 
participating   in the    trigger  with   either  NN-2  or NN-3 trigger configuration  and  using  a  
total  of 225 pixels. 
At present, the  telescope has a 5$\sigma$ sensitivity of detecting Crab Nebula in 25 hours of 
observation time.
This sensitivity  figure  has  been  verified  several  times  over  last  5 years  and  consolidated 
results  on the  Crab nebula,   for  $\sim$310.5 h   of observation time  taken over  a  period  of  3 
observing   seasons,  is in  preparation.. 
\par
For   validating  the Monte Carlo simulation  results  for   NN-2  trigger  mode  we  have  used   data  
on the  Crab Nebula  for $\sim$101.44 h between November 10, 2005 - January 30, 2006. In   this data sample, 
the number of $\gamma$-ray events  obtained after applying  the    Dynamic Supercuts   are  found   to be   
$\sim$(928$\pm$100) with a statistical significance of 
$\sim$9.40$\sigma$ [31].  Similarly, for   validating  the  simulation  results  for   NN-3  trigger  mode   
we  have  used  data  on the  Crab Nebula   for $\sim$105.16 h between November 03, 2007 - March  8, 2008. 
After applying  the    Dynamic Supercuts  to  this data  sample, the number of $\gamma$-ray events     are  
found   to be   $\sim$(968$\pm$90) with a statistical significance of $\sim$11.1$\sigma$.  Data  collected 
in the  NN-2 trigger mode,     during  November 10, 2005 - January 30, 2006,  has  also been   used  by us  
previously for  checking  the  validity of the ANN-based energy estimation procedure  and  for  determining  
the energy  spectrum   of  the  Crab  Nebula  in  the  energy range  1-16 TeV [31].  The $\gamma$-ray 
differential  spectrum  obtained   after  using  appropriate values of  effective collection area and 
$\gamma$-ray acceptance  efficiency  (along with  their  energy and zenith angle dependence)  was  found to  
be  follow  a    power law   $(d\Phi/dE=f_0 E^{-\Gamma})$  
with  $f_0=(3.12\pm0.48)\times 10^{-11} cm^{-2}s^{-1}TeV^{-1}$  
and $\Gamma=2.69\pm0.14$ [31]. 
While   the    details  of  the  standard Dynamic Supercuts  procedure  used  by  us   to separate 
$\gamma$-ray  like images from the  background cosmic rays  can   be  found  in [27,31], in  this   work,   
we  will  only  concentrate   on the  cosmic  ray  data  collected  by  the telescope  for  validating  the 
Monte Carlo simulation  estimates   regarding    effective collection area, threshold  energy  and  integral 
rates.   
\par
Starting  with  the  analysis of  the  NN-2  trigger  mode  data   collected  for $\sim$101.44 h;  we have  
first divided   the total  data into 4 spells, where each spell  corresponds to one  lunation 
period.   Without   correcting  for  inter-pixel  gain variation  and   without  applying  other  data  
selection cuts,  Fig.12a  shows  measured  values  of  cosmic-ray  rate  as a function of  zenith  angle.   
The zenith angle  plotted  in this  figure  has  been  multiplied  with  the  sign of  hour angle  so 
that   pre-upper transit and  post-upper transit  rates  can  be distinguished easily.  Each  point  in  
this   plot  represents  a  10  minute  observation  run   and  the  event rate  has been  calculated  
by dividing   the  number of events  recorded   in 600s.  The  behaviour  of the   measured  values  
of  cosmic-ray  rate  as a function of  zenith  angle   for the   NN-3  trigger   mode  is  shown  
in   Fig.12b. 
\begin{figure}[h]
\includegraphics*[width=0.9\textwidth,angle=0,clip]{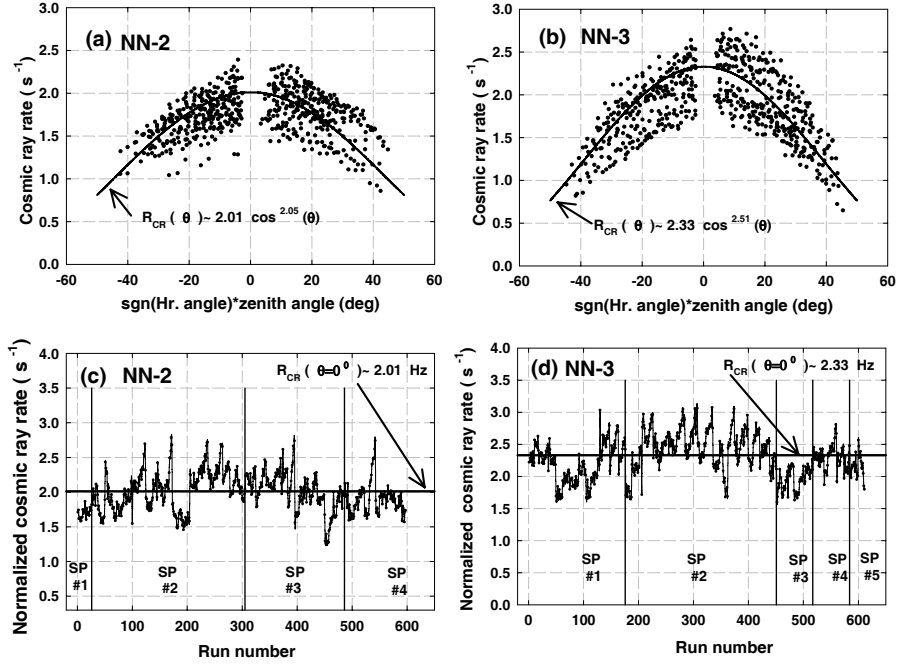}
\caption{ Measured  cosmic  ray  trigger  rate as a function of  zenith angle  (a) data  collected  between 
November 10, 2005 - January 30, 2006  with  NN-2 trigger configuration;(b) data  collected  between  
November 03, 2007 - March  8, 2008 with  NN-3 trigger configuration. The spline curve    shown  in theses 
figure   is  the  Monte Carlo-based  estimated   cosmic-ray trigger  rate. 
Normalized  cosmic ray rate  as  a function of  run number  (c)   November 2005 - January 2006  data with   
NN-2 trigger
 (d)   November 2007 - March  2008  data with   NN-3 trigger.  The  segmentation  of  the   data  into  
various  spells  are   indicated  by  full vertical lines.  Each spell corresponds to one  lunation period  
and   the  notation  SP$\#$  stands  for  spell number.} 
\end{figure}
In this case, the total data  has been  divided into 5 spells.  The  corresponding   Monte Carlo  
estimates (solid line), with  $R_{CR}(\theta)\sim2.012 (cos\theta)^{2.054} Hz$  for  NN-2 trigger  
and $R_{CR}(\theta)\sim2.330(cos\theta)^{2.510} Hz$ for NN-3 trigger,  are  also  shown   in   
Fig.12a  and  Fig.12b, 
respectively. Although  the  Monte Carlo  estimates  are  found   to  describe  the  zenith angle dependence 
of the measured  rates  reasonably  well   with  an  accuracy  of  $\pm \sim$20$\%$  for  most  of the 
10 minute runs,  presence  of a  systematic  uncertainty  in  the  experimental  data  can not  be ruled  
out  on account  of  sky condition variations, 
 both  on daily basis  as well as  from  one lunation to other. 
Differences  in  parameters  like  single  pixel trigger threshold,   overall photoelectron conversion 
factor etc.  are also  some  of the  additional factors  because  of which  there  can be   a  possible    
mismatch  between  simulated trigger   rates and  the corresponding experimental  values.   While work on  
understanding the telescope systematics is still in progress,  we  have   used  a   value   $ \sim15 \%$   
for   systematic error in  the  measured cosmic-ray rate  for  further  analysis.  The  justification for  
this     is  based on the guidance   from [32]  where  the  author   has   reported a  variation  
of  $\sim15 \% -20 \%$  in the  Chrenkov  photon  density  for  mid-latitude  sites  on 
account  of  seasonal  changes. Although,
one  does  not  expect  a  significant   variation   in  atmospheric  extinction  on a nightly basis for 
a   good  astronomical  site,  significant   variation   in  atmospheric  extinction  on  a nightly 
basis  at  Mt.  Abu   has  been  observed  by  us  over   the last several  years.  Using  a   value   
$ \sim15 \%$   for  systematic error in  the  measured cosmic-rate   is   thus  a fairly  conservative  
estimate.  
\par 
The measured  cosmic-ray rates  have also   been   normalized   to a  zenith angle of  0$^o$ and variation  
of  the  
normalized  rates  as  a  function  of  run  number  is   shown  in  Fig.12c (NN-2 trigger ) and  Fig.12d 
(NN-3 trigger).
In order   to quantify  departures   from   a   flat   shape,  we have  also calculated the  $\chi^2$  
by using   the following expression [33]
\begin{equation}
\chi^2 =  \sum_{i=1} ^{n_{run}} \frac{ (R_{CR,i} - R_{CR}^{MC})^2 } 
{  (\Delta R_{CR,i})^2 + (f_i ^{rel} R_{CR,i})^2 }\
\end{equation}
where $n_{run}$ is   the number of 10 minute  runs, $R_{CR,i}$  is   the    normalized measured  
cosmic-ray  rate   with 
$\Delta R_{CR,i}$  as  the   corresponding  statistical  error,  $f_i ^{rel}$  is the  relative  
systematic  uncertainty
and $R_{CR}^{MC}$ is  the normalized  Monte  Carlo  simulated trigger   rate.  The  resulting   values   
of   $\chi^2/dof \sim  605.4/ 597$ (probability = 0.40)  obtained  for   the  NN-2  trigger  observation  
spell   and   
$\chi^2/dof \sim  662.6 /611$ (probability = 0.07) obtained  for   the  NN-3  trigger  observation  spell  
clearly   suggest    
that  there  is    a   reasonably   good  matching    between  the   simulated trigger   rates and  the 
experimentally  measured  trigger   rates.  Presence of a systematic error   at    $>15\%$  level    for     
November 2007 - 
March 2008 data  may  also  be  contributing  to the  relatively  higher  value  of   $\chi^2$   for this  
data  spell.   
\par
The    homogeneity  of  the   triggers produced  within the trigger  field of view   is  another  test   
to  check   the   performance  of the trigger  generation  scheme.   
Although  finding the   frequency   distribution   of the   triggered  pixels  would   have   been   the  
best  way   to   validate  this, unavailability  of the  hit  pattern  data  at  the hardware  level  does  
not  allow  us to  check   this  directly.  To   circumvent  this,  we  determine  the   frequency   
distribution  of   the  image  centroid  coordinates  by using the following approach. First,  the data  
is  corrected for inter-pixel gain variation and  then subjected to the standard two-level image  
'cleaning' procedure [27] with picture and boundary thresholds of  6.5$\sigma$ and 3.0$\sigma$ 
respectively.    In the  clean  Cherenkov  image,  if  $ith$  PMT   with  coordinates  $x_i$, $y_i$  
registers a  signal  $s_i$,    the  coordinates   of the  image  centroid
are given by :  $<x>$=  $\sum_i s_i x_i$/ $\sum_i s_i$ and  $<y>$=  $\sum_i s_i y_i$/ $\sum_i s_i$.
 Considering  the   central $\sim$ $2.25^o$ $\times$ $2.25^o$  trigger  region and  dividing  it  into  
225  cells  of  size  $\sim$ $0.15^o$ $\times$ $0.15^o$  each,   we  bin   the  image  centroid  
co-ordinates  into   these  angular   cells.  Assuming   that  the image  centroid    is  very close  to 
the brightest  pixel (  i.e. pixel  with   highest   signal),   the  resulting   centroid  map   can  be  
faithfully used  for  checking the  homogeneity  of  the   triggers produced  within the trigger  field of 
view.   Defining   
 $ N_{cell,i}$  as   number of  events   in the  ith  cell  and  $ <N_{cell}>$  as  the average   number 
of events  in each  cell   we   calculate  percentage   residual  of   events ($\Delta N _{cell,i}$) in  
each  cell   where 
 $\Delta  N _{cell,i}$ =$ (N_{cell,i}$ -$<N_{cell}>)$/$<N_{cell}>$.
 In order  to ensure  that  images  are  robust and  are contained   within  the camera  we have  also 
used   the following  cuts  
 for  selecting  the  events  for this analysis :$\mid<x>\mid <1.2^o$,  $\mid<y>\mid<1.2^o$,  
$SIZE\geq 450 d.c$  where (6.5 digital counts $\equiv$1.0 pe ) and  number  of pixels  participating  in 
the  image $\geq4$.
Fig.13  shows  the   results  of  the    percentage  residual  events  
 ($\Delta  N _{cell,i}$)   for   the  NN-2    and  NN-3      observation  spells . 
\begin{figure}[h]
\includegraphics*[width=0.9\textwidth,angle=0,clip]{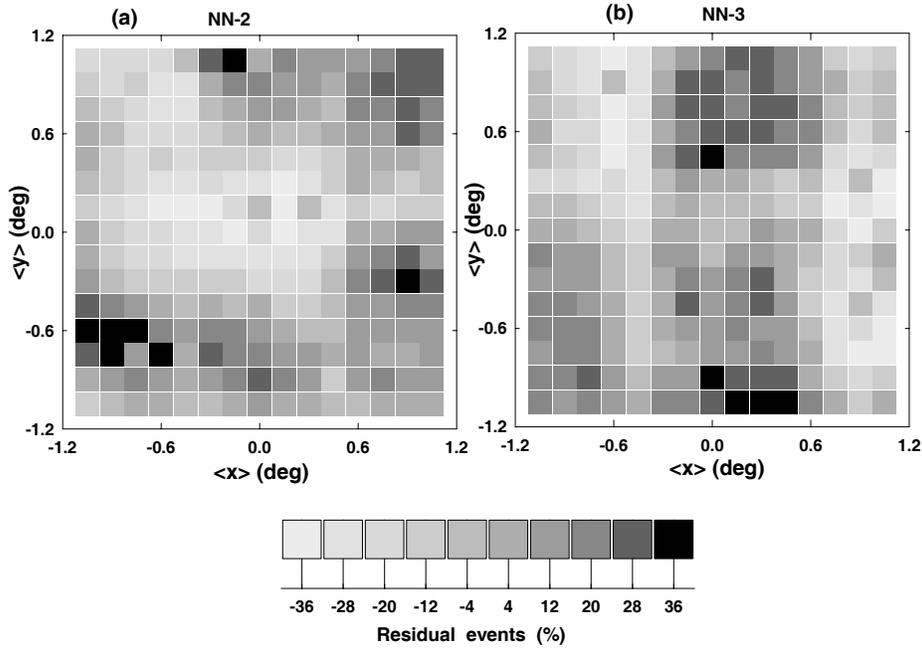}
\caption{ (a)  Percentage  residual  events  ( $\Delta  N _{cell,i}$ ), for   the  NN-2  data,   on the 
basis  of the  image  centroid  coordinates  by  following   gray  shade coding scheme. (b) Same as  (a) 
except   with  NN-3  trigger  configuration. The  gray  shade coding scheme  followed  for  representing   
the  percentage  residual  events  is  also  given  at the bottom of this  figure.}
\end{figure}
A   gray  shade coding scheme   has   been  used   in  this   figure  for  representing   the  percentage  
residual  events.  The  corresponding   results  obtained  for  the  NN-3  trigger  configuration   by 
using  data   collected  during   November 03, 2007 - March  8, 2008  is  shown  in  Fig.13b. In the   
NN-2  data ( Fig.13a)   we  find  that   $\sim 69\%$   of the cells   have  
$\mid \Delta  N _{cell,i}\mid$  with in   $\sim 20\%$.    The corresponding   number of  cells,  for   the  
NN-3 data ( Fig.13b),   is  found   to    be  $\sim 65\%$.  Presence  of   non-functional   pixels  in the  
camera (typically   in the range   of 3 to 7  in a total of  225   pixels   and   whose  location  keeps  
on  changing with time)   as  well as   excessive    trigger  rate  in  some  pixels  due to presence  of  
relatively  bright  stars   are  the   2 main   causes   because    of   which       
$\mid \Delta  N _{cell,i}\mid$  is   more   than    $\sim 30\%$   in   some regions  of the  camera.  
\par
 Compatibility of the  trigger  rates  with  Poissonian statistics   and   homogeneity  of  the   triggers 
produced  within the trigger  field of view   clearly suggest      that   the  overall   performance  of 
the  telescope is  quite  stable and  the  Monte Carlo simulation  estimates  regarding  effective 
collection area and $\gamma$-ray acceptance  are  also  quite  reliable. 
 The  agreement between the  expected and  actual  performance  of the telescope  has  also been  checked  
by  comparing the    expected  and observed  image parameter distributions   and  results   of this  
study  can be  seen in [10].
   The consistent detection  of  a steady signal from  the Crab Nebula  above 
$\sim$1.2 TeV energy, at a sensitivity level of 
$\sim$5.0 $\sigma$ in  $\sim$25 h,    alongwith  excellent matching of its energy spectrum  
with that obtained  by  other  groups,  also  reassures  that the  performance  of the TACTIC  telescope 
is  quite  stable and 
reliable  [27,29,31].
However,   more   work   has   to be done  with  regard  to  relative  calibration  of  the telescope  
( along  the lines  suggested  in [34])  so that   unavoidable systematic errors  as a   result  of  
temporal  changes  in the light collection efficiency, gain and,  most importantly, atmospheric   
conditions  can  be  accounted  for   in the analysis. 
\section{Conclusions}
\label{}
A   detailed    simulation  study   for  the TACTIC  telescope    has   been  presented  in this  work for  
optimizing   its   trigger  field  of  view  and   the  trigger  generation scheme.  Different   trigger  
regions  with  49,81,121,169 and  240  pixels  were  considered  in this  work  and  the  results   
suggest   that  a trigger  field  of   11$\times$11  ($\sim$ $3.4^o$ $\times$ $3.4^o$)  is quite  optimum   
for  achieving   maximum
detection rate   for   $\gamma$-rays  from a point  source.   With regard  to  optimization  of  
topological  trigger  generation,  it  is found  that   using   a  NN-2  trigger  methodology  is  a   
better  option  as  against   the   NN-3  trigger  configuration  when  a   source   needs to be  observed  
at   zenith angles of  $>$ 25$^o$.  
Reasonably good  matching   of the   measured  trigger  rates  (on  the basis   of  $\sim$ 207 hours of  
data  collected  with  the telescope  in  NN-2  and  NN-3   trigger configurations)   with  that obtained  
from simulations  reassures  that the procedure followed by us  in  estimating    the  threshold  energy 
 and  detection rates   is  quite  reliable.
Furthermore,  excellent matching of   Crab   Nebula   differential  spectrum    with that obtained  by 
 other  groups,  
reported by  us  in our  earlier   work,   also  reassures  that  the procedure followed by us for obtaining 
the energy spectrum  of a $\gamma$-ray  source is quite  reliable.   Attempts are also being   made at present 
to  improve  the  sensitivity   of  the telescope.  Apart  from  using  new  CPCs    ( modified   CPC  with    square  entry  and  a  circular  exit) for  enhancing  the  light  collection efficiency  of  the  camera,  we are  also  trying  to  reduce      the    coincidence  gate   width.   Both  these  modifications   are  expected  to   lower   the    threshold  energy of the   TACTIC telescope.  To conclude,  we  firmly   believe 
that there is  considerable scope  for a TACTIC  like imaging telescope for monitoring AGNs on a long
term basis.  

\section{Acknowledgements}
The authors would like to convey their gratitude to all the concerned colleagues of the  
Astrophysical Sciences  Division  for their contributions towards the instrumentation and 
observation aspects of the TACTIC telescope.  We would  also  like  to thank the anonymous  referee 
for  his valuable comments  and suggestions  which  have helped us to improve the quality  of the paper.  

\end{document}